\documentclass[12pt,a4paper]{article}

\usepackage{amsmath,amssymb} 
\usepackage{accents}
\usepackage{color}
\usepackage[bookmarks=true,hyperfigures=true]{hyperref}
\usepackage{graphicx}
\usepackage[nosort]{cite}
\usepackage[bulletsep]{collref}
\usepackage{tensor}
\usepackage{lscape}
\usepackage{empheq}

\usepackage{color}

\usepackage[a4paper,text={173mm,216mm},centering]{geometry}

\allowdisplaybreaks[3]

\usepackage[font=small,labelfont=bf,width=0.85\textwidth]{caption}

\let\oldbfseries=\bfseries
\let\oldmdseries=\mdseries
\let\oldnormalfont=\normalfont
\renewcommand{\bfseries}{\oldbfseries\boldmath}
\renewcommand{\mdseries}{\oldmdseries\unboldmath}
\renewcommand{\normalfont}{\oldnormalfont\unboldmath}

\makeatletter
\newlength{\apb@width}
\newcommand{\autoparbox}[2][c]{\settowidth{\apb@width}{#2}\parbox[#1]{\apb@width}{#2}}

\makeatother

\newcommand{\smallnumber}{\mathbf{\epsilon}}

\newcommand{\remark}[2][.]{{\color{red}\renewcommand{\bfdefault}{b}\rmfamily\if.#1\else\textbf{#1:} \fi#2}}

\DeclareMathOperator{\Pf}{Pf}                               
\newcommand{\be}{\begin{equation}}
\newcommand{\ee}{\end{equation}}
\newcommand{\beq}{\begin{equation}}
\newcommand{\eeq}{\end{equation}}
\newcommand{\bal}[1]{\setlength{\jot}{1em}\begin{align}\begin{aligned}#1\end{aligned}\end{align}}

\newcommand{\bma}{\begin{pmatrix}}
\newcommand{\ema}{\end{pmatrix}}
\newcommand{\ba}{\begin{eqnarray}}
\newcommand{\ea}{\end{eqnarray}}
\newcommand{\trans}{{\scriptscriptstyle\mathrm{T}}}         
 
\newcommand{\mR}{\mathcal{R}}
\newcommand{\mQ}{\mathcal{Q}}
\newcommand{\epr}{\tilde{\varepsilon}_{p,r}}

\ifx\genfrac\sdflkaj\else\fi

\newcommand{\cA}{\mathcal{A}}

\newcommand{\cO}{\mathcal{O}}

\def\l<{\langle}\def\r>{\rangle}

\makeatletter
\newcommand{\namedref}[2]{\hyperref[#2]{#1~\ref*{#2}}}
\newcommand{\secref}{\@ifstar{\namedref{Section}}{\namedref{sec.}}}
\newcommand{\subsecref}{\@ifstar{\namedref{Subsection}}{\namedref{subsec.}}}
\newcommand{\appref}{\@ifstar{\namedref{Appendix}}{\namedref{app.}}}
\newcommand{\tabref}{\@ifstar{\namedref{Table}}{\namedref{tab.}}}
\newcommand{\figref}{\@ifstar{\namedref{Figure}}{\namedref{fig.}}}
\makeatother

\newif\ifmrnote 
\mrnotetrue
 
\usepackage[active]{srcltx}

\newif\ifjbnote 
\jbnotetrue
 
\newcommand{\matr}[2]{\left(\begin{array}{#1}#2\end{array}\right)}

\def\[{\begin{equation}}
\def\]{\end{equation}}
\def\<{\begin{eqnarray}}
\def\>{\end{eqnarray}}

\newcommand{\vev}[1]{\langle#1\rangle}
\newcommand{\bev}[1]{ [#1]}

\newcommand{\eqn}[1]{(\ref{#1})}

\newcommand{\fb}{\bar{f}}

\def\bea{\begin{align}}
\def\eea{\end{align}}
\def\be{\begin{equation}}
\def\ee{\end{equation}}

\newcommand{\mP}[1]{\mathcal{P}_{#1}}  
\newcommand{\mJ}{\mathcal{J}}  
\newcommand{\pr}{(p\cdot r)}

\newcommand{\Cs}{\textbf{c}} 
\newcommand{\Si}{\textbf{s}} 
\newcommand{\csminus}{(\Cs^2-\Si^2)}

\begin{document}
\thispagestyle{empty}

\begingroup\raggedleft\footnotesize\ttfamily
HU-EP-16/26\\
\vspace{15mm}
\endgroup

\begin{center}
{\LARGE\bfseries Collinear limits beyond the leading order from the scattering equations
\par}%
\vspace{15mm}

\begingroup\scshape\large 
Dhritiman Nandan, Jan Plefka and Wadim Wormsbecher
\endgroup
\vspace{5mm}

\textit{Institut f\"ur Physik and IRIS Adlershof, Humboldt-Universit\"at zu Berlin, \phantom{$^\S$}\\
  Zum Gro{\ss}en Windkanal 6, D-12489 Berlin, Germany} \\[0.1cm]
\texttt{\small\{dhritiman.nandan,jan.plefka,wadim.wormsbecher\}@physik.hu-berlin.de\phantom{\ldots}} \vspace{8mm}


\textbf{Abstract}\vspace{5mm}\par
\begin{minipage}{14.7cm}
The structure of tree-level scattering amplitudes for collinear massless bosons is studied
beyond their leading splitting function behavior. These near-collinear limits at 
sub-leading order are best studied 
using the Cachazo-He-Yuan (CHY) formulation of the S-matrix based on the scattering equations. 
We compute the collinear limits for gluons, gravitons and scalars. It is shown that the CHY integrand for an $n$-particle gluon scattering amplitude in the collinear limit at sub-leading order
 is expressed as a convolution of an $(n-1)$-particle gluon integrand and a collinear kernel integrand, which is universal. Our representation is shown to obey
recently proposed amplitude relations in which the collinear gluons of same helicity
are replaced by a single graviton. Finally, we  extend our analysis to effective field theories and study the collinear limit of the non-linear sigma model, Einstein-Maxwell-Scalar and Yang-Mills-Scalar theory. 
\end{minipage}\par
\end{center}
\newpage



\setcounter{tocdepth}{2}

\tableofcontents
\section{Introduction}

A very important feature of scattering amplitudes is their factorization property
when generalized Mandelstam invariants go on-shell. The residues associated to such
propagator poles are given by the product of two lower-multiplicity amplitudes and this
analytic structure puts strong constraints on the functional form of the amplitudes. 
This property also lies at the heart of various modern on-shell techniques for
scattering amplitudes developed in the recent past being far more efficient than 
the traditional  Feynman diagram based approach\footnote{See \cite{Dixon:1996wi,Elvang:2013cua,Henn:2014yza} for a text-book level introduction.}.
Of particular interest are the factorization properties on the two-particle poles that 
involve three-point amplitudes which are subtle objects in massless quantum field theories.
For real momenta such a two-particle pole $s_{ij}=2p_{i}\cdot p_{j}=0$ implies collinearity
of the two momenta, $p_{i}=z\,p$ and $p_{j}=(1-z)p$ with $z\in[0,1]$. Color 
ordered gluon scattering amplitudes factorize in such a collinear limit according to 
\be
A_{n}(1^{h_{1}},2^{h_{2}},\ldots) \stackrel{1\parallel2}{\longrightarrow}
\sum_{h=\pm} \text{Split}_{-h}(z; 1^{h_{1}},2^{h_{2}})\,\, A_{n-1}(p^{h},\ldots),
\ee
with the universal splitting functions $\text{Split}_{-h}(z; 1^{h_{1}},2^{h_{2}})$, which 
diverge as $1/\sqrt{p_{1}\cdot p_{2}}$ and are known to very high loop-orders in QCD. Along with collinear properties other aspects of infra-red(IR) physics like behavior under soft limits also provide deep insights into nature. 
In the soft limit $p\to\delta\, q$ with $\delta\to 0$ color-ordered gluon amplitudes 
display a universal factorization\cite{Weinberg:1964ew,Low:1958sn,Burnett:1967km}
\be
A_{n+1}(\delta \, q, p_{1},..., p_{n}) \underset{\delta \to 0}{=}
(\frac{1}{\delta}S^{[{0}]}( q,\{p_{a}\})+S^{[1]}(q,\{p_{a}\})) \cdot A_{n}(p_{1},\ldots, p_{n})
+\cO({\delta}) \, .
\ee
Interestingly enough, this universal factorization extends beyond the leading singular
term in $\delta$, for gauge theories up to the sub-leading order\cite{Low:1958sn,Burnett:1967km,Casali:2014xpa}  and for gravity even to sub-sub-leading order\cite{Cachazo:2014fwa}. The universal factorization properties of gluon amplitudes in soft and
collinear limits have been studied extensively over many decades and lead to various
interesting properties including the all-loop universal form of IR divergences\cite{Bern:2005iz,Anastasiou:2003kj}. Moreover, recent insights into sub-leading and sub-sub-leading terms in the soft expansion for YM and gravity theories have revealed their connections to hidden symmetries of the asymptotic null infinity where soft theorems are manifest as Ward identities of such symmetries\cite{Strominger:2013jfa,He:2014laa,He:2015zea}.
In light of this interesting structure in the soft limit at sub-leading level it is a natural question to explore the collinear properties of scattering amplitudes at the sub-leading order as well. In this work we provide a framework for studying such scenarios in a systematic way for a wide class of quantum field theories and specifically focus on the explicit form for the collinear limit of tree-level amplitudes of gluons at the sub-leading order. In fact, computing such effects using standard textbook techniques is an arduous task requiring to take into account the sum over a large number of contributing Feynman diagrams in general. Modern on-shell methods usually provide a way to bypass these complications but unlike the sub-leading soft theorem a straightforward BCFW\cite{Britto:2004ap,Britto:2005fq} analysis of the gluon amplitudes with collinear kinematics up to the sub-leading level  suggests that all possible BCFW factorization channels contribute, thus making it very hard to find a compact and insightful expression.
%

However, a uniform formulation of tree-level scattering amplitudes for theories with massless particles in any dimension has recently been proposed by Cachazo, He and Yuan(CHY). The full tree-level S-matrix for the scattering of gluons and gravitons, as well as 
massless cubically interacting scalars, in arbitrary dimensions may be represented in an
universal fashion as an integral over the moduli space of a punctured Riemann sphere 
\cite{Cachazo:2013hca,Cachazo:2013iea}. The key ingredient are a set of
scattering equations
\be
f_a = \sum_{b=1 \atop b\not=a}^{n} \frac{2\,p_a \cdot p_b}{\sigma_a - \sigma_b} =0  \; ,
\label{eq-scateq}
\ee
where the $p_{a}^{\mu}$ denote the null-momenta of the scattered particles and the
$\sigma_{a}\in\mathbb{C}$ are the positions of the punctures. These equations have appeared a number
of times in the literature in various contexts\cite{Fairlie:2008dg,Gross:1987ar, Witten:2004cp, Caputa:2011zk, Caputa:2012pi, Makeenko:2011dm,Cachazo:2012uq}. They are known to possess $(n-3)!$ solutions for an $n$ particle scattering problem. This formulation of scattering amplitudes makes many dimension independent properties and symmetries manifest. Especially the study of single\cite{Cachazo:2013hca,Cachazo:2013iea} and multi-soft theorems \cite{Cachazo:2015ksa,He:2016vfi,Saha:2016kjr}\footnote{Also see \cite{Klose:2015xoa,Volovich:2015yoa} for other approaches to multi-soft theorems in field theory.} becomes quite straightforward. Inspired by the successful study of soft theorems in CHY we will utilize the same formalism to study collinear limits beyond the leading order.

On a parallel track of recent research, various intriguing relations have been observed between amplitudes in Yang-Mills theories and theories containing both gluons and gravitons, namely the Einstein-Yang-Mills (EYM) theory. These have been explored both from Type I\cite{Stieberger:2009hq,Stieberger:2015vya,Stieberger:2016lng} and heterotic string theory\cite{Schlotterer:2016cxa} as well as from field theory perspective\cite{Stieberger:2016lng,Nandan:2016pya,delaCruz:2016gnm}. A particularly interesting set of such identities holding in
the collinear limit were proposed by Stieberger and Taylor 
\cite{Stieberger:2015kia,Stieberger:2014cea}. Here specific linear combinations of $n$-gluon sub-leading collinear amplitudes in YM theory are related to amplitudes in EYM with one graviton and $(n-1)$ gluons. Effectively the two collinear gluons in the pure YM amplitude are being replaced by a graviton.  There is a nice understanding of this property from superstring theory \cite{Stieberger:2014cea}, but here a direct derivation in the field theory will
be performed. While intriguing the Stieberger-Taylor relations only provide half of the needed $(n-3)!$ relations to completely determine the sub-leading collinear
gluon structure. It is also interesting to note that some of the above mentioned relations in the quantum field theory regime find a natural description in terms of the CHY formalism\cite{Nandan:2016pya,delaCruz:2016gnm}. Since the CHY formula for various theories shares many common building blocks it is very suitable for studying relations among theories. One of the important motives for our framework to compute sub-leading collinear effects in quantum field theories is also to explore and prove such relations.  

In this work we report on a detailed study of the collinear limit of adjacent gluons in YM theory. Our main result is the amplitude for same helicity gluons $ 1 $ and $ 2 $ becoming collinear at
the sub-leading order in a near collinear limit in the form of  a CHY representation,
\be
\cA_{n,\,(1\parallel 2)}^{\text{YM},(1)}=\int d\mu_{n-1}\mathcal{K}^{\text{gluon}}_{\text{coll}}(\epsilon_p, \{p_i\},\{\sigma_i\})\, 
\mathcal{I}_{n-1}^{\text{Yang-Mills}}[(\epsilon_p;p), \{(\epsilon_{3;}p_3)\ldots 
(\epsilon_{n}; p_{n})\},\{\sigma_i\}].
\label{subleadingcollYMform}
\ee
Note that the above sub-leading CHY formula of the collinear YM $n-$gluon amplitude is given by the CHY integral over the  $(n-1)-$ particle YM integrand weighted by a collinear kernel $\mathcal{K}^{\text{gluon}}_{\text{coll}}$, whose form is given later in \eqref{eq:gluonkernel}, such that this kernel is only dependent on the helicity of the effective collinear particle $ \epsilon_p $ even though it depends on the momenta and the position of punctures for all the other particles. Hence, even though this collinear kernel at the sub-leading order cannot be taken outside the CHY integral to yield factorization in the form of a sub-leading splitting
function, it does have  robust universal structure that it is not sensitive to the type of the non-collinear particles. We also note the intriguing relation that a gauge transformation on
the effective leg $p$ of the sub-leading collinear amplitude yields an $(n-1)$-point
gluon amplitude
\be
p\cdot \frac{\partial}{\partial \epsilon_{p}}\cA_{n,\,(1\parallel 2)}^{\text{YM},(1)}=
\frac{2z-1}{z(z-1)}\, \cA_{n-1}^{\text{YM}}(p,3,\ldots,n)
\ee
Moreover, one can use the expression \eqn{subleadingcollYMform} to recover the Stieberger-Taylor relations within field theory such that this collinear kernel in fact magically combines to generate the CHY integrand building block for the scattering of one graviton with the remaining $(n-1)$-gluons in EYM theory.

As an extension to the new structures seen in the study of soft theorems for different theories it is important to mention that recently there has also been a lot of interest in exploring scattering amplitudes in certain effective field theories (EFT) which have been proposed to have compact CHY formulae \cite{Cachazo:2014xea} as well. The interest in these theories not only stems from their appearance in certain physical contexts but also due to the special behavior of their amplitudes under soft limits which could be used to classify such EFTs\cite{Cheung:2014dqa,Cheung:2015ota}. Not much is known about their properties under collinear limits. Their CHY formulation is very suitable for such a study due to the framework we develop here. In this work we also comment on the collinear limits of a number of EFTs.  

We organize the paper in the following way:
In section two we begin with setting-up the kinematics for the collinear limit that we will study. We then discuss solutions of the scattering equations and the behavior of various building blocks in the CHY formula under this limit. In section three we compute the leading order collinear limits of gluons and gravitons recovering the known gluonic and scalar splitting functions and find their graviton cousins. In section four we present the results of the sub-leading collinear limits for scalars and gluons with same helicity and the universal nature of the collinear CHY integrand and follow it up in section six with further explorations in universal structure by adding a soft limit on top of this.. In section six we use the sub-leading YM results and show that it directly reproduces the Stieberger-Taylor relations. Next in section seven, we extend our study to scattering amplitudes in EFTs, namely the Non-Linear Sigma Models(NLSM), YM-Scalar(YMS) and Einstein-Maxwell-Scalar(EMS) theories. We conclude with a summary and outlook in section eight. Finally, we have an appendix to present some of the details of the sub-leading collinear gluon computation and other observations that we made during this project.

\section{Collinear kinematics in the scattering equations}
In this section we introduce the reader to the basic concepts of the CHY formalism and we establish all necessary preliminaries for the computation of the collinear limits. 
\subsection{CHY formula}

In terms of the CHY formula \cite{Cachazo:2014nsa,Cachazo:2014xea,Cachazo:2015ksa} the scattering amplitude for $n$ massless bosons with momenta $p_{a}$ and 
polarizations $\varepsilon_{a}$ takes the general form of an integral over positions of the
punctures $ \sigma_a $ on a Riemann sphere
\be
\label{CHYmaster}
  \mathcal{A}_{n}(\{p,\varepsilon\}) = \int d\mu_{n}\, \mathcal{I}_{n}(\{p,\varepsilon,\sigma\}) \; ,
\ee
where 
\beq
d\,\mu_n := d'\sigma_n\,\Delta_n'= (\sigma_{ij}\,\sigma_{jk}\,\sigma_{ki})\,(\sigma_{pq}\,\sigma_{qr}\,\sigma_{rp})\prod_{a=1\atop{a\neq i,j,k}}^{n}\,d\sigma_a\,\prod_{b=1\atop b\neq{p,q,r}}^{n}\delta(f_b)~~, ~~ \sigma_{ab}:=\sigma_{a}-\sigma_{b}\label{eq:measure}~.
\eeq
We write $d'\sigma_n$ for the $\mathbb{CP}_1$ measure and $\Delta_n'$ for the product of 
$\delta$-functions which impose the scattering equations \eqref{eq-scateq}. The prime is the commonly used notation for deleting three arbitrary integrals and $\delta$-functions, manifest in the second equality. In this work we will focus on theories whose integrand,  $\mathcal{I}_{n}(\{p,\varepsilon,\sigma\})$, is constructed from a combination of two building blocks, namely the Park-Taylor factor  $ \mathfrak C_{n}(1,\ldots,n) $ and the Pfaffian of an anti-symmetric $2n\times 2n$ matrix $\Psi_n$, i.e.~ $\text{Pf}'\, \Psi_{n}(\{p,\varepsilon,\sigma\})$ where
\be
\mathfrak C_{n}= \frac{1}{\sigma_{12}\sigma_{23}\ldots \sigma_{n1}} \quad,\quad 
\Psi_{n} = \matr{cc}{A & -C^\trans \\ C & B}~,
\label{psimatrixparketaylor}
\ee
with the entries
\be
\label{ABCmatrices}
  A_{ab} = \begin{cases}
    \frac{2p_a\cdot p_b}{\sigma_a - \sigma_b} & a\not=b \; , \\
    0                                        & a=b\, , 
  \end{cases}
  \qquad
  B_{ab}^{h_a|h_b} = \begin{cases}
    \frac{2\varepsilon_a^{h_a}\cdot \varepsilon_b^{h_b}}{\sigma_a - \sigma_b} & a\not=b \; , \\
    0                                        & a=b\, , 
  \end{cases}
  \qquad
  C_{ab}^{h_a} = \begin{cases}
    \frac{2\varepsilon_a^{h_a}\cdot p_b}{\sigma_a - \sigma_b} & a\not=b \; , \\
    - \sum\limits_{c\not=a} \frac{2\varepsilon_a^{h_a}\cdot p_c}{\sigma_a - \sigma_c} & a=b\, .
  \end{cases}
\ee
The superscript $h_i$ is labeling the helicity state of particle $i$ and the prime of the Pfaffian indicates that two rows and columns must be deleted according to $\Pf'\, \Psi_{n}=\frac{(-1)^{i+j}}{\sigma_{ij}}\Pf\Psi_n^{i,j}$, where we use the superscript to denote which rows and columns $\{i,j\}\in \{1,...,n\}$ are deleted. Combining both building blocks one defines the tree-level $n$-particle S-matrix \cite{Cachazo:2013hca,Cachazo:2013iea} of pure gravity, pure Yang-Mills and scalar $\phi^{3}$-theory as
\bal{
\mathcal{I}_{n}^{\text{gravity}}= (\Pf'\,\Psi_{n})^{2} \quad,\quad\mathcal{I}_{n}^{\text{Yang-Mills}}= \mathfrak C_{n}\, \Pf'\,\Psi_{n} \quad,\quad
\mathcal{I}_{n}^{\text{scalar}}=\mathfrak C_{n}^{2}\, ~.
\label{eq:chyallintegrand}
}
Furthermore, single-trace S-matrices of Einstein-Yang-Mills theory\cite{Cachazo:2014nsa} can be constructed easily, e.g.~for $k$ gravitons and $n-k$ gluons we have
\beq
\mathcal I_n^{\text{EYM}}(\{p,\varepsilon,\sigma\})=\mathfrak C_{n-k}\,\Pf\Psi_k \, \Pf'\Psi_n. \label{eq:EYMint}
\eeq
Note that the Pfaffian $\Pf\Psi_k$ is not primed and that one should treat the indices of the integrands as sets, i.e.~the index $k$ on $\Psi_k$ means that only the graviton labels are present and $\mathfrak C_{n-k}$ only includes the labels of the gluons.

\subsection{The adjacent collinear limit in CHY} \label{The Collinear Limit in CHY}
Now we move on to discuss the proper implementation of the collinear kinematics in the CHY formalism. 
\subsubsection{Kinematic definitions}
Consider the scattering of $n$ particles thereby taking the neighboring particles $1$ and $2$ to be
 collinear. Then
their spinor-helicity variables\footnote{See e.g.~\cite{Dixon:1996wi,Elvang:2013cua,Henn:2014yza} for a text-book introduction.} may be conveniently parametrized as\cite{Stieberger:2015kia} 
\bal{
\label{colllimitdef}
|1\rangle &= \Cs\, |p\rangle -\smallnumber\, \Si\, |r\rangle \quad,\quad
|1] = \Cs\, |p] -\smallnumber\, \Si\, |r]\quad,\quad \Cs=\cos\phi~, \\
|2\rangle &= \Si\, |p\rangle +\smallnumber\, \Cs\, |r\rangle \quad,\quad
|2] = \Si\, |p] +\smallnumber\, \Cs\, |r] \quad,\quad \,\Si=\sin\phi~,
}
where we use the perturbation parameter $\smallnumber$ to probe the sub-leading collinear effect as $\smallnumber\to 0$.
Here $p^{\mu}=p_{1}^{\mu}+p_{2}^{\mu}+ \cO(\smallnumber^{2})$ is the limiting collinear momentum vector
and $r^{\mu}$ is a null reference momentum not parallel to $p^{\mu}$. 
This translates the four-momenta of the collinear particles to
\bal{
p_{1}&= \Cs^{2}\, p - \smallnumber\, \Cs\Si\, (|p\rangle [r| + |r\rangle [p|) + \smallnumber^{2} \, \Si^{2}\,r~,
\\
p_{2}&= \Si^{2}\, p + \smallnumber\, \Cs\Si\, (|p\rangle [r| + |r\rangle [p|) + \smallnumber^{2} \, \Cs^{2}\,r\label{eq:momentadef}~.
}
Defining $q=|p\rangle [r| + |r\rangle [p|$ we have $q^{2}=-2p\cdot r$ and
\be
(p_{1}+p_{2})=p+\smallnumber^{2}\,r\quad , \quad (p_{1}-p_{2})=(\Cs^2-\Si^2)(p-\smallnumber^{2}\,r)
-2\smallnumber\, \Cs\Si\, q\quad,\quad (p_{1}\cdot p_{2})= \smallnumber^{2}\, p\cdot r\, . 
\ee
Collinearity of two particles affects their polarizations. In four dimensions we have
\bal{
	\varepsilon^{+}_{a}= +\sqrt{2}\, \frac{|a]\, \langle r |}{\langle r \, a\rangle} \quad,\quad
	\varepsilon^{-}_{a}= -\sqrt{2}\, \frac{|a\rangle\, [ r |}{[ r \, a]}\, ~, 
}
where we have chosen the same $r$ as in \eqref{colllimitdef} to be the reference vector for all particles. Then
\bal{
	\varepsilon^{\pm}_{1} = \varepsilon^{\pm}_{p}-\smallnumber\, \frac{\Si}{\Cs}\, \epr^{\pm}
	\, \quad,\quad 
	\varepsilon^{\pm}_{2} = \varepsilon^{\pm}_{p}+\smallnumber\, \frac{\Cs}{\Si}\, \epr^{\pm}\, ~.
}
Here $\varepsilon_{p}^{\pm}$ is the polarization of a leg carrying the limiting 
collinear momentum $p$ and 
\be
\epr^{\pm} :=
\begin{cases} 
	+ \sqrt{2}\,\frac{|r\rangle\, [r|}{\langle r \, p \rangle} & \text{pos. helicity} \\
	- \sqrt{2}\,\frac{|r\rangle\, [r|}{[r  \,p ]}  & \text{neg. helicity}~.\\
\end{cases} 
\ee
A very important consequential identity is
\beq
\epr^{h_p}\cdot p = -\varepsilon_p^{h_p}\cdot q ~,
\eeq
which follows directly from $\varepsilon_1^{h_1}\cdot p_1 = \varepsilon_2^{h_2}\cdot p_2 = 0$. Although we are using spinor helicity notation here, we stress that all our results are straightforwardly lifted to general dimensions.

\subsubsection{Collinear scattering equations}
Next, we will analyze the scattering equations \eqref{eq-scateq} in collinear kinematics \eqref{colllimitdef}. Following the strategy of the double soft limit in \cite{Cachazo:2015ksa}, we perform a change of variables for the positions of the punctures 
corresponding to the two collinear momenta $ p_1 $ and $ p_2 $, i.e.
\beq
\sigma_{1}=\rho-\frac{\xi}{2} \quad,\quad \sigma_{2}=\rho+\frac{\xi}{2}~.
\label{collinearsigmas}
\eeq
Now we solve for $\xi$ by rewriting the measure as
 \bal{
 	d\mu_{n}&=d\mu_{n-2}\, d\Omega\\
 	d\Omega=
 	d\sigma_{1}\, d\sigma_{2}\, \delta(f_{1})\, \delta(f_{2})&=
 	2\, d\rho\, d\xi\, \delta(f_+)\, \delta(f_{-})\, ,\label{eq:deltatransform}
 }
 with 
 \be
 f_+ =(f_{1}+f_{2}) \,\quad , \quad  f_- = (f_{1}-f_{2})-\csminus(f_{1}+f_{2})\,~.
 \label{SEcombocoll}
 \ee
The choice of this peculiar linear combination for $f_-$ will be commented upon at the end of this section. Note that $d\Omega$ is never affected by the ``prime" operation defined in \eqref{eq:measure}.
The scattering equations now take the form
\bal{
\label{SEgeneric}
f_{a}&= \sum_{b=3 \atop b\not=a}^{n} \frac{2\,p_a \cdot p_b}{\sigma_a - \sigma_b}+\frac{2\,p_{a}\cdot p_{1}}{\sigma_{a}-\rho+\frac{\xi}{2}} +
\frac{2\,p_{a}\cdot p_{2}}{\sigma_{a}-\rho-\frac{\xi}{2}}\, , \qquad a\neq 1, 2\\
f_{1}-f_{2}&= \sum_{b=3}^{n} \Bigl (\frac{2\,p_{b}\cdot p_{1}}{\rho-\frac{\xi}{2}-\sigma_{b}} - \frac{2\,p_{b}\cdot p_{2}}{\rho+\frac{\xi}{2}-\sigma_{b}} \Bigr ) 
- 4\, \frac{p_{1}\cdot p_{2}}{\xi}\,  \\
f_{1}+f_{2}&= \sum_{b=3}^{n} \Bigl (\frac{2\,p_{b}\cdot p_{1}}{\rho-\frac{\xi}{2}-\sigma_{b}} + \frac{2\,p_{b}\cdot p_{2}}{\rho+\frac{\xi}{2}-\sigma_{b}} \Bigr ) 
\, .
}
 An interesting way of seeing the relation between the collinear kinematics $p_{1}||p_{2}$ of the two particles and the coincidence of $\sigma_{1}$ and $\sigma_{2}$ was already argued in \cite{Dolan:2014ega}. If we assume a linear vanishing of $\xi$ in the $\smallnumber\to 0$ limit \eqref{degansatz}, the scattering equations \eqref{SEgeneric} factorize at leading order in $\smallnumber \rightarrow 0$
 into a set of $(n-1)$-particle scattering equations
 for each of the $ (n-1) $  momenta $\{p,p_{3},\ldots, p_{n}\}$ 
 \begin{align}
 0=f_{a}\Bigr|_{\smallnumber\to 0}&= \sum_{b=3 \atop b\not=a}^{n} \frac{2\,p_a \cdot p_b}{\sigma_a - \sigma_b}+\frac{2\,p_{a}\cdot p}{\sigma_{a}-\rho} + \cO(\smallnumber)\, , \qquad
 0=f_{1}+f_{2}\Bigr|_{\smallnumber\to 0}= \sum_{b=3}^{n} 
 \frac{2\,p \cdot p_{b}}{\rho-\sigma_{b}}  + \cO(\smallnumber)\, .
 \end{align}
 along with  
 \be
 0=f_{1}-f_{2} \Bigr|_{\smallnumber\to 0} =- \frac{4}{\smallnumber}\, 
 \frac{p_{1}\cdot p_{2}}{\xi_{1}}\, \Bigl ( 1 + \cO(\smallnumber)\, \Bigr)\, \quad \Rightarrow
 \quad p_{1}\cdot p_{2}=0\, .
 \ee
 Hence, we see that coinciding $\sigma_{1}$ and $\sigma_{2}$
 implies collinearity of the corresponding particle momenta. The question then arises whether the opposite is also true, i.e.~whether collinearity implies confluency of $\sigma_{1}$ and $\sigma_{2}$.
To clarify this, we performed a numerical analysis using the polynomial form of the scattering equations\cite{Dolan:2014ega}. Dialing various kinematical configurations with two collinear momenta for $n=5,6,7$ and $8$ particles we have consistently found two different classes of solutions in all cases tested, namely there are  $2(n-4)!$ degenerate solutions with vanishing $ \xi =\sigma_{21}$, accompanied by $(n-5)(n-4)!$ non-degenerate solutions with finite $\xi$. In fact, the above-mentioned numerics also support a vanishing of $\xi$ linear in $\smallnumber$ and not with a higher power. We also observe that the non-degenerate solutions only contribute to the limiting amplitudes at the sub-sub-leading order in the $\smallnumber$-expansion. Hence for all our analysis in this work where we only focus up to the sub-leading order we can exclusively consider the degenerate solutions captured by the ansatz 
\be
\xi = \smallnumber\, \xi_{1}+\smallnumber^{2}\, \xi_{2}+  \cO(\smallnumber^{3})\, ~.
\label{degansatz}
\ee
We insert this ansatz into the scattering equations obtaining
\bal{
	f_{a}&=\fb_a+\smallnumber\biggl[-\csminus\frac{\xi_1}{2}\mathcal{P}_{2,a}\biggr ]+\cO(\smallnumber^{2})\, , \qquad a\neq 1, 2\\
	f_+&= \fb_p+\smallnumber\biggl[ (\Cs^2-\Si^2)\frac{\xi_1}{2} \mP2  \biggr ]+\cO(\smallnumber^{2}) \\
	f_{-}&=\smallnumber\left [2\Cs^2\Si^2\xi_1\mP2-2\Cs\Si\mQ_1 -\frac{4\,\pr}{\xi_1}\right ]
	\\
	& \quad+\smallnumber^2\left[ 4\,\pr \frac{\xi_2}{\xi_1^2} -2(\Cs^2-\Si^2)\mR_1 +2\Cs^2\Si^2\xi_2\mP2 +\xi_1(\Cs^2-\Si^2)\Cs\Si\mQ_2 \right] +\cO(\smallnumber^{3})\, ,
	\label{newlistfexp}
}
where we have introduced the modified scattering equations for $ n-1 $ particles as\footnote{For $\fb_a$ the sum starts from $0$ which indicates the momentum of the effective collinear particle $p_0=p$ and the full sum runs over the set of momenta$\{p,p_3,p_4,\ldots, p_n\}.$},
\be
\fb_a=\sum_{b=0 \atop b\not=a}^{n} \frac{2\,p_a \cdot p_b}{\sigma_a - \sigma_b}  \quad, \quad
\fb_p=\sum_{b=3 }^{n} \frac{2\,p \cdot p_b}{\rho - \sigma_b} \quad,\quad3\leq a\leq n+2~, 
\ee
along with the short-hand notations,
\bal{\label{SEcurlysymbols}
	&\mathcal P_{i,a}=\frac{2\,p\cdot p_a}{(\rho - \sigma_a)^i} ~~,~~\mP i = \sum_{b=3 }^{n} \frac{2\,p \cdot p_b}{(\rho - \sigma_b)^{i}} ~~ , ~~ \mathcal{R}_{i}= \sum_{b=3 }^{n} \frac{2\,r \cdot p_b}{(\rho - \sigma_b)^{i}}
	\\&\mathcal{R}_{i,a}= \frac{2\,r \cdot p_a}{(\rho - \sigma_b)^{i}}~~,~~
	\mQ_{i} = \sum_{b=3 }^{n} \frac{2\,q \cdot p_b}{(\rho - \sigma_b)^{i}} ~~,~~ \mQ_{i,a} = \frac{2\,q\cdot p_a}{(\rho-\sigma_b)^i}~~,~~i\geq 1 \, ,
}
for compactness. Note that $\mP 1=\fb_{p}=\cO(\smallnumber)$ by virtue of the scattering-equation corresponding to the particle with momentum $ p $. Also we now see why it is advantageous to consider the specific linear combination $f_-$ in \eqref{SEcombocoll} as it is $\mathcal O(\smallnumber)$.
%
\subsubsection{The degenerate collinear solutions}
In order to find the degenerate solutions of $\xi$ to the scattering equations in the collinear limit \eqref{newlistfexp} we perform the $\xi$ integral by localizing the $\delta$-function
$\delta(f_-)$. This amounts to solving  the equations $ f_- =0$ in \eqref{newlistfexp} perturbatively in 
$\smallnumber$. At the leading order in $\cO(\smallnumber)$  we have
\be
-2\,\pr-\Cs\Si\mQ_1\xi_1+\Cs^2\Si^2\xi_1^2\mP2=0 \,,
\label{ziOep1eqn}
\ee
with solutions
\bal{
\xi_1 &=\xi_{1}^\pm=  x_1\pm \sqrt{x_2} \quad, \\
x_1 &= \frac{\mQ_1}{2\Cs\Si\mP2}\quad, \quad  x_2=\frac{\mQ_1^2+8\pr \mP2}{ 4(\Cs^2\Si^2)\mP2^2} \, .
\label{zisolnforX}
}
Moving on to the next order at $\cO(\smallnumber^2)$, it can be easily established that 
\be
\frac{\xi_{2}}{\xi_{1}^{2}}= \frac{\Cs^{2}-\Si^{2}}{2} \, \frac{2\mR_{1}-\Cs\Si\,\xi_1\, \mQ_{2}}{2\,p\cdot r + \Cs^{2}\Si^{2}\, \xi_{1}^{2}\, \mathcal{P}_{2}} \,~.
\label{xi2solution}
\ee
Note the vanishing of the sub-leading solution $\xi_{2}$ in the $\Cs = \Si$ case.

\subsection{Collinear expansion of the CHY formula}\label{BBstudy}
 Inserting the transformations for the collinear $ \sigma $'s \eqref{collinearsigmas}, the measure \eqref{eq:deltatransform} and the collinear kinematics \eqref{colllimitdef} into the CHY formula \eqref{CHYmaster} one obtains
\bal{
	\mathcal A_n = 2\,\int d\mu_{n-2}\,d\rho\,d\xi\,\delta(f_{+})\,\delta(f_{-})\,\mathcal I_n = 2\,\sum_{\xi_1} \,\int\,d\mu_{n-2}\,d\rho \,\delta(f_{+}) \,\mathcal J\,\mathcal I_n~,   
\label{CHYmastercollinear}}
where we performed the $\xi$ integral by using
\be
\label{deltafminusJ}
\delta(f_{-})=\sum_{\xi^\pm}\frac{\delta(\xi-\xi^\pm)}
{\quad\left |\frac{\partial f_{-}}{\partial\xi}\right|_{\xi=\xi^\pm}} =\sum_{\xi^\pm}\delta(\xi-\xi^\pm)\,\mathcal J
 ~,
\ee
such that $\mJ$ is the Jacobian for the transformation $\xi\to f_{-}$ and the sum runs over the two solutions \eqref{ziOep1eqn} for $\xi$. In order to complete our analysis we need the expansion of all the different components that make up the CHY formula.
\paragraph{Jacobian.}   
We write the expansion of the Jacobian in \eqref{deltafminusJ} as
\bal{
\mJ&= \mJ_0 + \smallnumber\,\mJ_1 + \mathcal O(\smallnumber^2)~,}
with
\bal{
&\mJ_0= \frac{1}{2}\,\frac{\xi_{1}^{2}}{2\pr+\Cs^2\Si^2\mP2\, \xi^{2}_{1}}~,\\  &\mJ_1=\mJ_0^2\,\biggl( 8\pr\frac{\xi_2}{\xi_1^3}-\Cs\Si(\Cs^2-\Si^2)\mQ_2\biggr)~.
\label{jacobianexpOep}
}
Note that, using this notation, we can also write \eqref{xi2solution} as,
\be
\xi_2 = \csminus\mJ_0 \left(2\mR_1 - \Cs\Si\, \xi_1\, \mQ_2 \right)~.\label{xi2withJac}
\ee
\paragraph{Measure.}
The measure in \eqref{CHYmastercollinear} can be expressed as,
\be
d\mu_{n-2}\,d\rho \,\delta(f_{+}) = d\mu_{n-1}^{(0)} + \smallnumber\,d\mu_{n-1}^{(1)} + \mathcal O(\smallnumber^2)~,
\ee
with
\be
d\mu_{n-1}^{(0)} = d\mu_{n-2}\,d\rho\,\delta(\bar f_p)~,
\ee
\be
d\mu_{n-1}^{(1)} = (\Cs^2 - \Si^2)\, d'\sigma_{n-2}\,d\rho\,\frac{\xi_1}{2}\,\left(\mathcal P_2 \,\delta'(\bar f_p)\Delta'^{(0)}_{n-2}- \delta(\bar f_p)\sideset{}{'}\sum_{a=3}^{n} \mathcal P_{2,a} \delta'(\bar f_a)\,\Delta'^{(0)}_{n-3,a}\right)~,
\ee
where 
\bal{
&\delta'(x)=\frac{\partial\,\delta(x)}{\partial\, x}\quad,\quad\Delta'^{(0)}_{n-3,a}=\prod_{i=3\atop i\neq a}^n \delta(\bar f_a)\quad,\quad
\sideset{}{'}\sum_{a=3}^{n}=(\sigma_{ij}\,\sigma_{jk}\,\sigma_{ki})\,\sum_{a=3 \atop  a\neq i,j,k}^n~.}
The definition of the primed sum is due to \eqref{eq:measure}.
\paragraph{Parke-Taylor factor.} 
Here we easily obtain,
\bal{
&\mathfrak C_{n}=\frac{1}{\smallnumber}\mathfrak C_{n}^{(0)}+\mathfrak C_{n}^{(1)}+\mathcal O(\smallnumber)~,\\
&\mathfrak C_{n}^{(0)}=-\frac{1}{\xi_1}\mathfrak C_{n-1}\,\quad ,\quad\mathfrak C_{n}^{(1)}=\mathfrak C_{n-1}\,\biggl(\frac{\xi_2}{\xi_1^2} +\frac{1}{2}\,S_{n,\rho,3}\biggr)~,\\
&\mathfrak C_{n-1}=\frac{1}{\sigma_{34}\ldots \sigma_{n\rho}\, \sigma_{\rho 3}} \quad ,\quad S_{i,j,k}= \frac{\sigma_{ik}}{\sigma_{ij}\,\sigma_{jk}} = \frac{1}{\sigma_{ij}} + \frac{1}{\sigma_{jk}}~.
\label{PT-linear}
}
%
\paragraph{The matrix $\Psi_{n}$.}
We write the matrix \eqref{psimatrixparketaylor} with more emphasis on the entries which are sensitive to the collinear limit as
\be
\Psi_{n}=
\begin{pmatrix}
	0 & A_{12} & A_{1b} &
	-C_{11}^{h_1} 
	&
	-C_{21}^{h_2}
	&
	-C_{d1}^{h_d} \\
	A_{21} & 0 & A_{2b} &
	-C_{12}^{h_1} &
	-C_{22}^{h_2} &
	-C_{d2}^{h_d} \\
	A_{a1} &  A_{a2} & A_{ab} & -C_{1a}^{h_1} & -C_{2a}^{h_2} & -C_{da}^{h_d} \\
	C_{11}^{h_1} &
	C_{12}^{h_1} &
	C_{1b}^{h_1} & 
	0 & B_{12}^{h_1|h_2} & B_{1d}^{h_1|h_d} \\ 
	C_{21}^{h_2} &
	C_{22}^{h_2} &
	C_{2b}^{h_2}&
	B_{21}^{h_2|h_1} & 0 & B_{2d}^{h_2|h_d} \\
	C_{c1}^{h_c}& C_{c2}^{h_c} & C_{cb}^{h_c} &
	B_{c1}^{h_c|h_1} & B_{c2}^{h_c|h_2} & B_{cd}^{h_c|h_d} \\
\end{pmatrix}~.
\ee
We now expand out all entries by using the definitions from section \ref{The Collinear Limit in CHY} and list them below (modulo higher orders in $\smallnumber$):
\begin{enumerate}
\item \textbf{$A$-matrix:}	
  \bal{
  	&A_{1b}=\Cs^2\,A_{pb} + \smallnumber\left(\frac{\xi_1}{2}\,\Cs^2\,A_{pb}^{(2)}- c\,s\,Q_{b}\right)~, 
  	~A_{2b}=\Si^2\,A_{pb} - \smallnumber\left(\frac{\xi_1}{2}\,\Si^2\,A_{pb}^{(2)}- c\,s\,Q_{b}\right)~,\\
  	&A_{12}=-\smallnumber\frac{2\,p\cdot r}{\xi_1} + \smallnumber^2\,2\,p\cdot r\frac{\xi_2}{\xi_1^2}~. 
  }
\item \textbf{$B$-matrix:}  
  \bal{
  	&B_{1b}^{h_1|h_b}=B_{pb}^{h_1|h_b} + \smallnumber\,\frac{\xi_1}{2}\,B_{pb}^{h_1|h_b,(2)}  \quad,\quad B_{2b}^{h_2|h_b}=B_{pb}^{h_2|h_b} - \smallnumber\,\frac{\xi_1}{2}\,B_{pb}^{h_2|h_b,(2)} ~, \\
  	&B_{12}^{h_1|h_2}= \begin{cases}
  		0 & h_1\,=\,h_2 \\
  		\frac{2}{\smallnumber\,\xi_1} - \frac{2\,\xi_2}{\xi_1^2}  & h_1\,\neq\, h_2
  	\end{cases}~.
  }
\item \textbf{$C$-matrix:}  
  \bal{
  	&C_{b1}^{h_b}=\Cs^2\,C_{bp}^{h_b}-\smallnumber\,\left(\frac{\xi_1}{2}\,\Cs^2\,C_{bp}^{h_b,(2)} + \Cs\Si\,C_{bq}^{h_b}\right) ~,~
  	C_{b2}^{h_b}=\Si^2\,C_{bp}^{h_b}-\smallnumber\,\left(\frac{\xi_1}{2}\,\Si^2\,C_{bp}^{h_b,(2)} - \Cs\Si\,C_{bq}^{h_b}\right) ~,\\
  	&C_{1b}^{h_1}=C_{pb}^{h_1}+\smallnumber\left(\frac{\xi_1}{2}C_{pb}^{h_1,(2)}-\frac{\Si}{\Cs}\,E_b^{h_1}\right)~~,~~
  	C_{2b}^{h_2}=C_{pb}^{h_2}+\smallnumber\left(-\frac{\xi_1}{2}C_{pb}^{h_2,(2)}+\frac{\Cs}{\Si}\,E_b^{h_2}\right)~,\\
  	&C_{12}^{h_1}=\frac{\Si}{\Cs}\,\frac{2\,\epr^{h_1}\cdot p}{\xi_1} - \smallnumber\,\frac{\Si}{\Cs}\,\frac{2\,\epr^{h_1}\cdot p}{\xi_1}\,\frac{\xi_2}{\xi_1}~~,~~
  	C_{21}^{h_2}=\frac{\Cs}{\Si}\,\frac{2\,\epr^{h_2}\cdot p}{\xi_1} - \smallnumber\,\frac{\Cs}{\Si}\,\frac{2\,\epr^{h_2}\cdot p}{\xi_1}\,\frac{\xi_2}{\xi_1}~, \\
  	&C_{11}^{h_1} = C_{pp}^{h_1} - \frac{\Si}{\Cs}\,\frac{2\,\epr^{h_1}\cdot p}{\xi_1} + \smallnumber\,\left(\frac{\xi_1}{2}\,C_{pp}^{h_1,(2)}+\frac{\Si}{\Cs}\,\frac{2\,\epr^{h_1}\cdot p}{\xi_1}\,\frac{\xi_2}{\xi_1} +\frac{\Si}{\Cs}\,E^{h_1}\right)~, \\
  	&C_{22}^{h_2} = C_{pp}^{h_2} - \frac{\Cs}{\Si}\,\frac{2\,\epr^{h_2}\cdot p}{\xi_1} + \smallnumber\,\left(-\frac{\xi_1}{2}\,C_{pp}^{h_2,(2)}+\frac{\Cs}{\Si}\,\frac{2\,\epr^{h_2}\cdot p}{\xi_1}\,\frac{\xi_2}{\xi_1} -\frac{\Cs}{\Si}\,E^{h_2}\right) ~,
  	\\
  	&C_{aa}^{h_a} = C_{aa}^{h_a} + \smallnumber\,(\Cs^2-\Si^2)\,\frac{\xi_1}{2}\,C_{ap}^{h_a,(2)}~.
  }
\end{enumerate}
In the above we defined the short hand notations
 \bal{
 	&C_{bq}^{h_b}=\frac{2\,\varepsilon_{b}^{h_b}\cdot q}{\sigma_{b}-\rho}\,\quad , \quad A_{pb}^{(i)}=\frac{2\,p\cdot p_b}{(\rho-\sigma_b)^i}\, \quad , \quad B_{pb}^{h_p|h_b,(i)}=\frac{2\,\varepsilon_p^{h_p}\cdot\varepsilon_b^{h_b}}{(\rho - \sigma_b)^i}~,
 	\\&C_{bp}^{h_b,(i)}=\frac{2\,\varepsilon_b^{h_b}\cdot p}{(\sigma_b - \rho)^i}\, \quad ,\quad C_{pb}^{h_p,(i)}=\frac{2\,\varepsilon_p^{h_p}\cdot p_b}{(\rho-\sigma_b)^i} \quad , \quad C_{pp}^{h_p,(i)} = - \sum_{b=3}^{n}\frac{2\,\varepsilon_p^{h_p} \cdot p_b}{(\rho - \sigma_b)^i}~,\\
 		&E_b^{h_r} = \frac{2\,\epr^{h_r}\cdot p_b}{\sigma_{\rho b}}\quad,\quad E^{h_r}= \sum_{b=3}^{n} E_b^{h_r}
 		 ~.
 }
\paragraph{The Pfaffian $\Pf'\Psi_n$.}
Now we will study the expansion of the Pfaffian building block for two different cases, i.e.~whether the collinear bosons have equal or opposite helicity. The Pfaffian of $\Psi_n$ is computed via the recursive formula 
\be
\Pf'\Psi_n= \frac{(-1)^{n+1}}{\sqrt 2}\,\sum_{j=1\atop j\neq i}^{2n} (-1)^{i+j+1+\theta(j-i)}\, \Psi_{ij}\, 
\Pf'\Psi_n^{i,j}~,
\label{pfaffiannormalized}
\ee
where we introduced an appropriate normalization for future convenience. We expand along row $j$, $\Theta(j-i)$ is the Heaviside function, $\Psi_{i,j}$ is the matrix element at position $(i,j)$ and $\Psi_n^{i,j}$ is the matrix $\Psi_n$ with both deleted rows and columns $i,j$. By definition $\Pf(\Psi_n^{1,...,n}) = 1$. It is possible to add a multiple of a row and its corresponding column to another row and corresponding column without changing the value of the Pfaffian. The derivative of the Pfaffian may be written as
\be
\frac{\partial}{\partial \smallnumber} \Pf '\Psi_n = \frac{(-1)^{n+1}}{\sqrt 2}\,\sum_{i=1}^{2n-1}\sum_{j=i}^{2n}(-1)^{i+j+1} \frac{\partial \Psi_{i,j}}{\partial \smallnumber} \Pf' \Psi_n^{i,j}~.\label{eq:PfaffianDerviative}
\ee
Finally note that any operation performed on the Pfaffian always acts on both, rows and columns.
\begin{enumerate}
	\item \textbf{Equal helicity case.}
This case is characterized by $h_1=h_2=h$ and since we use the same reference vector for all polarizations we have $B_{12}^{h_1|h_2} = 0$. In anticipation of the sub-leading order, we perform the following manipulations:  
	\begin{enumerate}\label{manipulation}
		\item Add row/column 1 to 2.
		\item Subtract $\Cs^2$ times the new second row/column from the first row/column.
	\end{enumerate}
	Doing so, we end up with
	\be
	\Psi_{n}=
	\begin{pmatrix}
		0 & A_{12} & \smallnumber\,\tilde A_{1b} &
		-\Si^2\,C_{11}^{h} + \Cs^2\,C_{12}^h 
		&
		\Cs^2\, C_{22}^h -\Si^2\,C_{21}^h
		&
		-\smallnumber\,\tilde C_{d1}^{h_d} \\
		A_{21} & 0 & \tilde A_{2b} &
		-C_{12}^h-C_{11}^h &
		-C_{22}^h-C_{21}^h &
		-\tilde C_{d2}^{h_d} \\
		\smallnumber\,\tilde A_{a1} &  \tilde A_{a2} & A_{ab} & -C_{1a}^h & -C_{2a}^h & -C_{da}^{h_d} \\
		\Si^2\,C_{11}^h-\Cs^2\,C_{12}^h &
		C_{12}^h+C_{11}^h &
		C_{1b}^h & 
		0 & 0 & B_{1d}^{h|h_d} \\ 
		\Si^2\,C_{21}^h-\Cs^2 C_{22}^h &
		C_{22}^h+C_{21}^h &
		C_{2b}^h&
		0 & 0 & B_{2d}^{h|h_d} \\
		\smallnumber\,\tilde C_{c1}^{h_c} & \tilde C_{c2}^{h_c} & C_{cb}^{h_c} &
		B_{c1}^{h_c|h} & B_{c2}^{h_c|h} & B_{cd}^{h_c|h_d} \\
	\end{pmatrix}~,
	\label{Psiman}
	\ee
	with
	 \bal{
	 	& \tilde A_{1b} = \xi_1 \Si^2\Cs^{2}2 A_{pb}^{(2)}-\Cs\Si\mathcal Q_b + \mathcal O(\smallnumber)~,
	 	\\& \tilde A_{2b} = A_{pb} + \smallnumber\,(\Cs^2-\Si^2)\,\frac{\xi_1}{2} A_{pb}^{(2)} + \mathcal O(\smallnumber^2)~,
	 	\\&\tilde C_{d1}^{h_d} = \xi_1 \Si^2\Cs^{2}C_{dp}^{h_d,(2)}+\Cs\Si C_{dq}^{h_d} + \mathcal O(\smallnumber)~,
	 	\\&\tilde C_{d2}^{h_d} = C_{dp}^{h_d} - \smallnumber\, (\Cs^2-\Si^2)\,\frac{\xi_1}{2} C_{dp}^{h_d,(2) }+ \mathcal O(\smallnumber^2)\label{eq:mentriesaftexp}	 	
	 ~.}
Note that $A_{12}$ is also of $\mathcal O (\smallnumber)$. We expand the Pfaffian along the first row/column and write
\beq
\Pf'\Psi_{n} = \Pf'^{(0)}\Psi_n + \smallnumber\,\Pf'^{(1)}\Psi_n + \mathcal O (\smallnumber^2)~.
\eeq
Immediately we see that the only contribution to the leading order is
\bal{
	\Pf'^{(0)}\Psi_n&=\frac{-1}{\sqrt 2}(\Si^2\,C_{11}^{h} - \Cs^2\,C_{12}^{h})\Pf'(\Psi_n^{1,n+1}) -\frac{1}{\sqrt 2}\,(\Cs^2\,C_{22}^{h} - \Si^2\,C_{21}^{h})\Pf'(\Psi_n^{1,n+2}) \\&= -\frac{1}{\sqrt 2}\,(C_{pp}^{h} - C_{12}^{h} - C_{21}^{h})\Pf'(\Psi_{n-1}) = -\frac{1}{\sqrt 2}\,\left(C_{pp}^{h} - \frac{2\,\epr^h\cdot p}{\Cs\,\Si\,\xi_1}\right)\,\Pf'\Psi_{n-1} ~,\label{eq:equalpf0}
} 
since all other contributions in the Pfaffian expansion are proportional to $\smallnumber$ due to our manipulation of the matrix $\Psi_n$. Furthermore, $\Psi_n$ with deleted first row/column will have the two equal rows/columns $(n+1)$ and $(n+2)$ in the collinear limit wherefore deleting rows/columns $(1)$ and $(n+1)$ will produce the same Pfaffian as the matrix with deleted rows/columns $(1)$ and $(n+2)$. This was used in the second equality and we call the resulting matrix $\Psi_{n-1}$ since it is the matrix associated with the $(n-1)$ particle amplitude where the first particle has momentum $p^\mu$ and position $\rho$ on $\mathbb{CP}_1$, i.e.
\be
\Psi_{n-1} = 
\begin{bmatrix}
	0 & A_{pb} &

	-C_{pp}^{h} &
	-C_{dp}^{h_d} \\
	A_{ap} & A_{ab} & -C_{pa}^{h} & -C_{da}^{h_d} \\ 
	C_{pp}^{h} &
	C_{pb}^{h}&
	0 & B_{pd}^{h|h_d} \\
	C_{cp}^{h_c} & C_{cb}^{h_c} &
	B_{cp}^{h_c|h} & B_{cd}^{h_c|h_d} \\
\end{bmatrix}\, .
\ee
The sub-leading structure $\Pf'\Psi_n^{(1)}$ is rather involved and given in the appendix.
	\item \textbf{Opposite helicity case.}
	In the situation that the collinear gluons have opposite helicity, e.g.~$1^+~,~ 2^-$, the matrix \eqref{Psiman} will not have the feature of containing two equal rows/columns in the limit $\smallnumber\rightarrow 0$. Even worse, it seems like the matrix is now carrying a divergence in the $B_{12} = \frac{1}{\smallnumber\,\xi_1}$ component. We attack both issues by denoting the matrix after doing the usual manipulation pattern as
	\be
	\Psi_{n}=
	\begin{pmatrix}
		0 & A_{12} & \smallnumber\,\tilde A_{1b} &
		-\Si^2\,C_{11}^{+} + \Cs^2\,C_{12}^+ 
		&
		\Cs^2\, C_{22}^- -\Si^2\,C_{21}^-
		&
		-\smallnumber\,\tilde C_{d1}^{h_d} \\
		A_{21} & 0 & \tilde A_{2b} &
		-C_{12}^+-C_{11}^+ &
		-C_{22}^--C_{21}^- &
		-\tilde C_{d2}^{h_d} \\
		\smallnumber\,\tilde A_{a1} &  \tilde A_{a2} & A_{ab} & -C_{1a}^+ & -C_{2a}^+ & -C_{da}^{h_d} \\
		\Si^2\,C_{11}^+-\Cs^2\,C_{12}^+ &
		C_{12}^++C_{11}^+ &
		C_{1b}^+ & 
		0 & B_{12}^{+|-} & B_{1d}^{+|h_d} \\ 
		\Si^2\,C_{21}^--\Cs^2 C_{22}^- &
		C_{22}^-+C_{21}^- &
		C_{2b}^-&
		B_{21}^{-|+} & 0 & B_{2d}^{-|h_d} \\
		\smallnumber\,\tilde C_{c1}^{h_c} &\tilde C_{c2}^{h_c} & C_{cb}^{h_c} &
		B_{c1}^{h_c|+} & B_{c2}^{h_c|-} & B_{cd}^{h_c|h_d} \\
	\end{pmatrix}~.
	\label{Psimanmix}
	\ee
	The crucial insight now is to perform two more matrix manipulations, i.e.~
	\begin{enumerate} \label{manipulation2}
		\item Add $\smallnumber \,\Cs\,\Si\,\epr^-\cdot p$ times the $(n+1)$'st row/column to the first row/column , 
		\item Add $\smallnumber \,\Cs\,\Si\,\epr^+\cdot p$ times the $(n+2)$'nd row/column to the first row/column ,
	\end{enumerate}
	while using the four dimensional identities (we are using the short hand notations \eqref{SEcurlysymbols})
	\bal{
		&\mathcal Q_{1,i} = (\epr^+\cdot p)\,\,C_{pi}^{-} + (\epr^-\cdot p)\,\,C_{pi}^{+}\quad,\quad \mathcal Q_1= -\,(\epr^+\cdot p)\,\,C_{pp}^{-} - \,(\epr^-\cdot p)\,\, C_{pp}^{+}~,
		\\&C_{iq}^{+}=(\epr^+\cdot p)\,\, B_{ip}^{h_i|-}\quad,\quad C_{iq}^{-}=(\epr^-\cdot p)\,\, B_{ip}^{h_i|+}
		~,}
	which can be checked explicitly. If we also use \eqref{ziOep1eqn} to express $A_{12}$ in terms of $\mathcal P_2 $ and $\mathcal Q$, we end up with a first row of the form (modulo higher order terms in $\smallnumber$)
	\be
	\begin{pmatrix}
		0 ,& -\smallnumber\,\xi_1\,\Cs^2\,\Si^2\,\mathcal P_2 ,& \smallnumber\xi_1 \Cs^2 \Si^2\, A_{pb}^{(2)} ,& -\Si^2C_{11}^{+} ,& \Cs^2 C_{22}^{-},& -\smallnumber\xi_1 \Cs^2 \Si^2 C_{bp}^{+,(2)}\label{eq:mixfirstline}
	\end{pmatrix}~.
	\ee
	Again we write
	\beq
	\Pf'\Psi_{n} = \Pf'^{(0)}\Psi_n + \smallnumber\,\Pf'^{(1)}\Psi_n + \mathcal O (\smallnumber^2)~,
	\eeq
	where now all terms contribute, since we have a $\frac{1}{\smallnumber}$ term in the matrix, i.e.
	\bal{
		&\Pf'\Psi_n^{(0)}=\frac{-1}{\sqrt 2}\,\left[\,\Si^2\,\left(C_{pp}^+ - C_{12}^+\right)\,\Pf'\Psi_n^{(1,n+1)} + \,\Cs^2\,\left(C_{pp}^- - C_{21}^-\right)\,\Pf'\Psi_n^{(1,n+2)}\,\right] +\mathcal G  
		\\&=\frac{-1}{\sqrt 2}\,\left[\,\Si^2\,\left(C_{pp}^+ - \frac{\Si}{\Cs}\frac{2\,\epr^+\cdot p}{\xi_1}\right)\,\Pf'\Psi_{n-1}^{-} + \,\Cs^2\,\left(C_{pp}^- - \frac{\Cs}{\Si}\frac{2\,\epr^-\cdot p}{\xi_1}\right)\,\Pf'\Psi_{n-1}^{+}\,\right] +\mathcal G ~.\label{eq:mixpf0}
		} 
	Here we defined $\Psi_{n-1}^- = \Psi_n^{(1,n+1)}$ , $\Psi_{n-1}^+ = \Psi_n^{(1,n+2)}$ , since those are the matrices for a $(n-1)$ particle scattering where the first particle  either has positive or negative helicity respectively. The object $\mathcal G$ contains all other terms of the expansion along the first row that, in the equal helicity case, have been of $\mathcal O(\smallnumber)$ but are now $\mathcal O(1)$ due to the singular behavior of $B_{12}$ which we extracted from the Pfaffian by expanding along row $(n+1)$ (w.r.t.~the Pfaffian with undeleted rows/columns). The important observation is that $\mathcal G$ is independent of $\xi_1$ due to $B_{12}\propto \frac{1}{\xi_1}$. Again, the sub-leading order of the Pfaffian is given in the appendix.
\end{enumerate} 
\subsection{Key sum-over-solutions identities}
In order to evaluate the leading and sub-leading behavior of any amplitude it is necessary to sum over the two solutions of $\xi$. Any building block will contribute with a certain power of $\xi_1$ or $\xi_2$ either in the numerator or the denominator but since $\xi_2$ and $\mathcal J_1$ can be both expressed through $\xi_1$ and $\mathcal J_0$, the calculations will always narrow down to the computation of 
\be
\sum_{\xi_1^\pm} \frac{\mathcal J_0^N}{\xi_1^M} = \left((x_1+\sqrt{x_2})^{N-M} +(-1)^N(x_1 - \sqrt{x_2})^{N-M}\right)\frac{1}{(2\,\sqrt{x_2}h_1)^N}~.\label{eq:calctool}
\ee
This equality follows from  
\be
\mathcal J_0\,\Big|_{\xi_1^\pm} =\frac{1}{2}\frac{1}{\underbrace{2\,\Si^2\,\Cs^2\,\mP2}_{h_1}+\underbrace{(-\Cs\Si\mQ_1)}_{h_2}\frac{1}{\xi_1^{\pm}}}= \frac{\xi_{1}^\pm}{h_1\,\xi_1^\pm + h_2}=\pm\frac{(x_1 \pm \sqrt{x_2})}{2\,\sqrt{x_2} h_1 }~,
\ee
where we used \eqref{ziOep1eqn} (for the readers convenience $x_1 = \frac{\mQ_1}{2\Cs\Si\mP2}$, $ x_2=\frac{\mQ_1^2+8\pr \mP2}{ 4(\Cs^2\Si^2)\mP2^2} $) and $h_1\,x_1 + h_2 = 0$. Using this identity we note the key sum-over-solutions identities  
 \bal{
 	&\sum_{\xi_1}\mathcal J_0\,\xi_1^2 = \frac{\mathcal Q_1^2 + 2\,(p\cdot r)\,\mathcal P_2}{4\,\Cs^4\,\Si^4\,\mathcal P_2^3}\quad,\quad\sum_{\xi_1}\mathcal J_0\,\xi_1 = \frac{\mathcal Q_1}{2\,\Cs^3\,\Si^3\,\mathcal P_2}\quad,\quad \sum_{\xi_1} \mathcal J_0 = \frac{1}{2\,\Cs^2\,\Si^2\mathcal P_2}~,
 	\\&\sum_{\xi_1} \frac{\mathcal J_0}{\xi_1}=0\quad ,\quad \sum_{\xi_1} \frac{\mathcal J_0}{\xi_1^2}=\frac{1}{4\,p\cdot r}\quad,\quad \sum_{\xi_1} \frac{\mathcal J_0}{\xi_1^3} = -\frac{\Cs\,\Si\,\mathcal Q_1}{8\,(p\cdot r)^2}~,
 	\\& \sum_{\xi_1} \frac{\mathcal J_0}{\xi_1^4}=\frac{\Cs^2\,\Si^2\,\left(\mathcal Q_1^2 + 2\,(p\cdot r)\,\mathcal P_2\right)}{16\,(p\cdot r)^3}\quad,\quad 
 	\sum_{\xi_1} \frac{\mathcal J_0^2}{\xi_1} = \frac{\mathcal Q_1}{4\,\Cs^3\,\Si^3\,D\,\mathcal P_2}\quad,\quad \sum_{\xi_1} \frac{\mathcal J_0^2}{\xi_1^2} = \frac{1}{2\,\Cs^2\,\Si^2\,D}~,
 	\\&\sum_{\xi_1} \frac{\mathcal J_0^2}{\xi_1^3}=-\frac{\mathcal Q_1}{8\,\Cs\,\Si\,p\cdot r\,D}\quad,\quad\sum_{\xi_1} \frac{\mathcal J_0^3}{\xi_1^4}=\frac{1}{16\,\Cs^2\,\Si^2\,p\cdot r\,D}\quad,\quad \sum_{\xi_1} \frac{\mathcal J_0^3}{\xi_1^5}=-\frac{\mathcal Q_1}{32\,\Cs\,\Si\,(p\cdot r)^2\,D}~ ,\label{eq:data1}}
 where we defined $D:= \mathcal Q_1^2 + 8\,p\cdot r\,\mathcal P_2$. Immediately more identities follow, i.e.
 \bal{
 	&\sum_{\xi_1} \frac{\mathcal J_0 \xi_2}{\xi_1^2} =  \frac{(\Cs^2-\Si^2)}{4\,\Cs^s\,\Si^2\,D}\left(4\,\mathcal R_1 - \frac{\mathcal Q_1\,\mathcal Q_2}{\mathcal P_2}\right)\quad,\quad\sum_{\xi_1} \frac{\mathcal J_0 \xi_2}{\xi_1^3} =-\frac{\Cs^2-\Si^2}{2\,\Cs\,\Si\,D}\,\left(\mathcal Q_2 + \frac{\mathcal Q_1\,\mathcal R_1}{2\,p\cdot r}\right)~,
 	\\&\sum_{\xi_1} \frac{\mathcal J_0^2 \xi_2}{\xi_1^4} = \frac{(\Cs^2-\Si^2)\mathcal R_1}{8\,\Cs^2\,\Si^2\,p\cdot r\,D}\quad,\quad\sum_{\xi_1} \frac{\mathcal J_0^2 \xi_2}{\xi_1^5}=-\frac{\Cs^2-\Si^2}{16\,\Cs\,\Si\,p\cdot r\,D}\left(\frac{\mathcal R_1\,\mathcal Q_1}{p\cdot r} +\mathcal Q_2\right) ~.\label{eq:data2}
 }
\section{Leading order collinear limit}
Now we are ready to compute the leading collinear structure for several theories. We write the expansion of the amplitude as,
\beq
\mathcal A_n^{1||2} = \mathcal A_n^{(0)} + \smallnumber\,\mathcal A_n^{(1)} + \mathcal O(\smallnumber^2)~,
\eeq
and will now study collinear gluons, gravitons and scalars correspondingly.
\subsection{Collinear gluons}
We expect to find the known universal behavior in terms of the helicity-dependent split function\footnote{See \cite{Bern:2004cz} and citations therein.}, i.e.
\beq
\mathcal A_{n}^{(0)}(1^{h_{1}},2^{h_{2}},\ldots) =
\sum_{h=\pm} \text{Split}^{\text{tree}}_{-h}(\Cs; 1^{h_{1}},2^{h_{2}})\,\, A_{n-1}(p^{h},\ldots) +\cO(\smallnumber^{0})~,
\label{gluoncollinearLOknown}
\eeq
with
\bal{
&\text{Split}^{\text{tree}}_{+}(\Cs;1^{+},2^{+}) =0 \quad,\quad
\text{Split}^{\text{tree}}_{+}(\Cs;1^{+},2^{-}) =-\frac{1}{\smallnumber}\, \frac{\Si^{3}}{\Cs\, \vev{pr}}~,\\
&\text{Split}^{\text{tree}}_{-}(\Cs;1^{+},2^{+}) =\frac{1}{\smallnumber}\, \frac{1}{\Cs\,\Si\, \vev{pr}}\quad,\quad \text{Split}^{\text{tree}}_{-}(\Cs;1^{+},2^{-}) =\frac{1}{\smallnumber}\, \frac{\Cs^{3}}{\Si\, \bev{pr}} ~.
}
\paragraph{Same helicities ($h_1 = h_2 = +$):}
Inserting the leading contributions of all building blocks from the previous section to \eqref{CHYmastercollinear} yields
\bal{
	\mathcal A_{n}^{(0)} &= \frac{-1}{\sqrt 2}\,2\,\sum_{\xi_1}\int d\mu_{n-1}^{(0)}\,\frac{-\mathcal J_0}{\smallnumber\,\xi_1}\,\mathfrak C_{n-1}\,\left(C_{pp}^+-\frac{2\,\epr^+\cdot p}{s\,c\,\xi_1}\right)\,\Pf'\Psi_{n-1}\\
	&=\frac{-1}{\sqrt 2}\,\frac{2\,\epr^+\cdot p}{\smallnumber\,s\,c\,2\,p\cdot r}\,\int d\mu_{n-1}^{(0)}\,\mathfrak  C_{n-1}\,\Pf'(\Psi_{n-1})= \frac{1}{\smallnumber\,c\,s\,\langle p \, r\rangle}\,\mathcal A_{n-1}\\
	&=\text{Split}_{-}^{\text{tree}}\,\mathcal A_{n-1}~,
}
which is the known collinear behavior. Here the sum-identity $\sum_{\xi_1}\,\frac{\mathcal J_0}{\xi_1} = 0$ of the last section was used. It is easy to see that this statement is also true for $h_1 = h_2=-$.
\paragraph{Opposite helicities ($h_1 = +~,~h_2=-$):}
Due to $\sum_{\xi_1}\,\frac{\mathcal J_0}{\xi_1} = 0$, the object $\mathcal G$ drops out of the calculation and we are left with 
\bal{
	\mathcal A_{n+2}^{(0)}(1^+,2^-,...)=\text{Split}_{-}^{\text{tree}}\mathcal A_{n-1}(P^+,...) + \text{Split}_{+}^{\text{tree}}\mathcal A_{n-1}(P^-,...
	) ~,}
which again reproduces the right behavior. 	In both cases above, the universal structure is independent of the helicities or the type of the non-collinear particles participating in the scattering process.
\subsection{Collinear gravitons}
From our previous analysis and the form of the gravity integrand in $\mathcal A_n=\int d\mu_{n}\, \mathcal{I}^{\text{gravity}}_{n}$ \eqref{eq:chyallintegrand} it is evident that, since gravitons appear without any Parke-Taylor structure in the CHY formalism, no singular behavior would be present. The collinear expansion is finite and can be easily computed. For the case of pure gravity and identical helicities we have
\beq
\mathcal A_n^{(0)} (1^{hh},2^{hh},\cdots)= \sum_{\xi_1}\int d\mu_{n-1}^{(0)}\,\mJ_0 \,\left(C_{pp}-\frac{2\,\epr^h \cdot p}{\Cs\,\Si\,\xi_1}\right)^2\,\Pf'\Psi_{n-1}\,\Pf'\Psi_{n-1}~,
\eeq
with the result
\beq
\mathcal A_n^{(0)} = \frac{(\epr^h\cdot p)^2}{\Cs^2\,\Si^2\,p\cdot r}\mathcal A_{n-1}(P^{hh},\cdots)+\frac{1}{2\,\Cs^2\,\Si^2}\int d\mu_{n-1}^{(0)}\,\frac{C_{pp}^2}{\mathcal P_2}\,\Pf'\Psi_{n-1}\,\Pf'\Psi_{n-1}~,\label{eq:gravL}
\eeq
which is a mixture of a factorized part, that agrees with known results\cite{Bern:1998xc,Bern:1998sv}, and a previously unknown non-factorized integral kernel part. This structure is independent of the type of all other particles in the scattering process as long as two gravitons are going collinear, i.e. it is universal. We can, for example, look at the scattering of $m$ gravitons and $k$ gluons, represented through
\beq
\mathcal A_n = \int d\mu_n \,\mathfrak  C_k\,\Pf\Psi_m\,\Pf'\Psi_n~,
\eeq
which in the collinear graviton limit will have the form
\beq
\mathcal A_n^{(0)} = \sum_{\xi_1}\int d\mu_{n-1}^{(0)}\,\mJ_0\,\mathfrak  C_k \,\left(C_{pp}-\frac{2\,\epr^h \cdot p}{\Cs\,\Si\,\xi_1}\right)^2\,\Pf\Psi_{m-1}\,\Pf'\Psi_{n-1}~,
\eeq
and hence
\beq
\mathcal A_n^{(0)} = \frac{(\epr^h\cdot p)^2}{\Cs^2\,\Si^2\,p\cdot r}\mathcal A_{n-1}+\frac{1}{2\,\Cs^2\,\Si^2}\int d\mu_{n-1}^{(0)}\,\mathfrak C_k\,\frac{C_{pp}^2}{\mathcal P_2}\,\Pf'\Psi_{m-1}\,\Pf'\Psi_{n-1}~.
\eeq
Adding scalars will not change this behavior either.
\subsection{Collinear scalars}
Inserting the collinear expansion of the building blocks in \eqref{eq:chyallintegrand} we arrive at
\bal{
	\mathcal A_n = \int d\mu_{n}\, \mathcal{I}^{\text{scalar}}_{n}\vert_{\text{coll}}=2\,\sum_{\xi_1}\,\int d\mu_{n-1}^{(0)} \mathcal J_0 \, \frac{1}{\smallnumber^2\,\xi_1^2}\, \mathfrak  C_{n-1}^2 ~.
}
The leading order is easily computed to be
\beq
\mathcal A_n^{(0)} = \frac{1}{\smallnumber^2 \,2\,p\cdot r}\,\mathcal A_{n-1}~,
\eeq
which is an expected result one can obtain from Feynman graph calculations, i.e.~it is the factorization into a 3-vertex and a propagator with the $n-1$ point amplitude. Again, the exchange of particle types of any other particles except the collinear scalars is not changing this behavior.
\section{Universal structure in the sub-leading collinear limit}
We now compute the sub-leading order for the case of collinear gluons and scalars. Other bosonic theories do not have any singular behavior. For gluons, we will only consider the equal helicity case since the mixed helicity case is a tedious, yet straightforward generalization of the
results presented here.
\subsection{Gluons with equal helicity}
The sub-leading order of the amplitude is formally given by the following terms
\bal{
	\mathcal A_{n}^{(1)} =2\,\sum_{\xi_1} &\int d'\sigma_{n-2}\,d\rho\Big[\Delta'^{(0)}_{n-1} \mathfrak  C_{n}^{(1)}\mathcal J_0 \Pf'^{(0)}\Psi_{n} +\,\Delta'^{(0)}_{n-1} \mathfrak  C_{n}^{(0)}\mathcal J_1 \Pf'^{(0)}\Psi_{n} \\&+ \Delta'^{(0)}_{n-1} \mathfrak  C_{n}^{(0)}\mathcal J_0 \Pf'^{(1)}\Psi_{n} + \,\Delta'^{(1)}_{n-1} \mathfrak  C_{n}^{(0)}\mathcal J_0 \Pf'^{(0)}\Psi_{n}\Big]~,
	\label{fullcollinearsubLexpansion}}
where we expanded out all building blocks to the next-to-leading order.
The computation of the four terms is rather involved and is relocated to the appendix. The result is
\bal{
	\mathcal A_{n}^{(1)} = -\frac{1}{\sqrt{2}}\,&\int d\mu_{n-1}^{(0)}\left(\frac{C_{pp}^h}{2\,\Si^2\,\Cs^2\,\mathcal P_2}S_{n,\rho,3} + \frac{\Cs^2 - \Si^2}{2\,\Cs^2\,\Si^2\,\mathcal P_2}\,C_{pp}^{h,(2)}\right)\,\mathfrak  C_{n-1}\,\Pf'\Psi_{n-1}\\
	&-\frac{1}{\sqrt{2}}\,\frac{\Cs^2-\Si^2}{2\,\Cs^2\,\Si^2}\,\int d\mu_{n-1}^{(0)}\,\frac{C_{pp}^h}{\mathcal P_2}\,\mathfrak  C_{n-1}\,\frac{\partial}{\partial \rho}\left(\Pf'\Psi_{n-1}\Delta'^{(0)}_{n-1}\right)\label{eq:subl1}~.
}
Using partial integration, we find
\bal{
	\mathcal A_{n}^{(1)} = -\frac{1}{\sqrt{2}}\,&\int d\mu_{n-1}^{(0)}\Bigg(\frac{C_{pp}^h}{2\,\Si^2\,\Cs^2\,\mathcal P_2}S_{n,\rho,3} + \frac{\Cs^2 - \Si^2}{\Cs^2\,\Si^2\,\mathcal P_2}\,\left(C_{pp}^{h,(2)}-\frac{C_{pp}^h\mathcal P_3}{\mathcal P_2}\right)\\&-\frac{\Cs^2-\Si^2}{2\,\Cs^2\,\Si^2}\,\frac{C_{pp}^h}{\mathcal P_2}\left(S_{n,\rho,3} - \frac{2}{\sigma_{\rho3}}\right)\Bigg)\,\mathfrak  C_{n-1}\,\Pf'\Psi_{n-1}\\
	&-\frac{1}{\sqrt{2}}\,\frac{\Cs^2-\Si^2}{2\,\Cs^2\,\Si^2}\,\int d'\sigma_{n-2}\,d\rho\,\frac{\partial}{\partial \rho}\left(\frac{C_{pp}^h}{\mathcal P_2}\,\mathfrak  C_{n-1}\,\Pf'\Psi_{n-1}\Delta'^{(0)}_{n-1}\right)~,
}
which may be rewritten directly as
\bal{
	\mathcal A_{n}^{(1)} = -\frac{1}{\sqrt{2}}\,&\int d\mu_{n-1}^{(0)}\Bigg(\frac{C_{pp}^h}{\mathcal P_2}\left(\frac{1}{\Cs^2}\,\frac{1}{\sigma_{n\rho}}+\frac{1}{\Si^2}\,\frac{1}{\sigma_{\rho3}}\right) + \frac{\Cs^2 - \Si^2}{\Cs^2\,\Si^2\,\mathcal P_2}\,\left(C_{pp}^{h,(2)}-\frac{C_{pp}^h\mathcal P_3}{\mathcal P_2}\right)\Bigg)\,\mathfrak  C_{n-1}\,\Pf'\Psi_{n-1}\\
	&-\frac{1}{\sqrt{2}}\,\frac{\Cs^2-\Si^2}{2\,\Cs^2\,\Si^2}\,\int d'\sigma_{n-2}\,d\rho\,\frac{\partial}{\partial \rho}\left(\frac{C_{pp}^h}{\mathcal P_2}\,\mathfrak  C_{n-1}\,\Pf'\Psi_{n-1}\Delta'^{(0)}_{n-1}\right)\label{eq:finalSubL}~.
}
The last term is a total derivative w.r.t $\rho$ and can be dropped. Furthermore, we define the collinear gluon kernel by
\beq
\mathcal K_{\text{coll}}^{\text{gluon}}(\varepsilon_p^h,\{p_i\},\{\sigma_i\}) =-\frac{1}{\sqrt{2}}\,\left( \frac{C_{pp}^h}{\mathcal P_2}\left(\frac{1}{\Cs^2}\,\frac{1}{\sigma_{n\rho}}+\frac{1}{\Si^2}\,\frac{1}{\sigma_{\rho3}}\right) + \frac{\Cs^2 - \Si^2}{\Cs^2\,\Si^2\,\mathcal P_2}\,\left(C_{pp}^{h,(2)}-\frac{C_{pp}\mathcal P_3}{\mathcal P_2}\right)\right)~,\label{eq:gluonkernel}
\eeq
s.t. we can write the final result compactly as 
\beq \boxed{
	\mathcal A_{n}^{(1)} = \int d\mu_{n-1}^{(0)}\,\mathcal K_{\text{coll}}^{\text{gluon}}(\varepsilon_p^h,\{p_i\},\{\sigma_i\})\,\mathcal{I}_{n-1}^{\text{Yang-Mills}}}~.\label{eq:fullfinal}
\eeq
We may also ask the question whether this expression is universal, i.e.~invariant under the exchange of all other particle types except the collinear gluons and their direct neighbors. The answer is yes, which we shall proof in the appendix. Therefore $\mathcal K_{\text{coll}}^{\text{gluon}}(\varepsilon_p^h,\{p_i\},\{\sigma_i\})$ is a universal object reflecting
a factorization at the level of CHY integrands. Furthermore, note that we can express the kernel in a way s.t. it contains CHY building blocks, i.e.
\bal{
\sqrt{2}\,\mathcal K_{\text{coll}}^{\text{gluon}}(\{p_i\},\{\sigma_i\})  &= \frac{1}{\Cs^2}\,\left(\frac{1}{2\,\mathcal P_2}\,\varepsilon_p^{\bar h}\,\frac{\partial}{\partial p_n} +\frac{1}{\sqrt{2}\,\Pf\Psi_p}\frac{\partial}{\partial \rho}\,\frac{1}{\mathcal P_2}\right)\,(\Pf\Psi_p)^2
\\&-\frac{1}{\Si^2}\,\left(\frac{1}{2\,\mathcal P_2}\,\varepsilon_p^{\bar h}\,\frac{\partial}{\partial p_3} +\frac{1}{\sqrt{2}\,\Pf\Psi_p}\frac{\partial}{\partial \rho}\,\frac{1}{\mathcal P_2}\right)\,(\Pf\Psi_p)^2 ~,
}
where we defined $\bar h$ to be the opposite helicity to $h$ and $\Psi_p = \begin{pmatrix}
0& - C_{pp} \\
C_{pp} & 0
\end{pmatrix}$, i.e.~the matrix for the data of one particle with momentum $p$ and helicity $h$ at position $\rho$. Note that the derivatives in this expression are not acting on $\mathcal{I}_{n-1}^{\text{Yang-Mills}}$. \\ 
Although \eqref{eq:fullfinal} is a compact expression, we have been unable to find any factorized structure on the integrated  amplitudes level from this result. The reason is mainly the appearance of $\mathcal P_2$ in the denominator. Even in the case $\Cs^2=\Si^2$, i.e. 
\beq
	\mathcal A_{n}^{(1)} = -\frac{2}{\sqrt{2}}\,\int d\mu_{n-1}^{(0)}\,\frac{C_{pp}}{\mathcal P_2}\,S_{n,\rho,3}\,\mathfrak  C_{n-1}\,\Pf'\Psi_{n-1}~,
	\eeq 
no factorization can be extracted. Note that the reason for non-factorizability is of the same type as already seen in the gravity result \eqref{eq:gravL}. 	


We close the discussions of the sub-leading collinear gluon limit with two important
observations: Firstly, from the form of our final result \eqn{eq:gluonkernel} and \eqn{eq:fullfinal}
it is manifest that the  subleading-collinear gluon amplitude $\mathcal A_{n}^{(1)}$ 
for the case of identical helicities of the collinear legs, is bi-linear in the effective polarization vector $\epsilon^{h}_{p}$. This is not at all
 obvious from the outset. Secondly, making a gauge transformation on this
effective `fused' leg $p$, i.e.~acting with $p\cdot\partial_{\epsilon^{h}_{p}}$, of $\mathcal A_{n}^{(1)}$,  yields an $(n-1)$-point gluon amplitude,
\be
p\cdot\frac{\partial}{\partial \epsilon^{h}_{p}}\, \mathcal A_{n}^{(1)}
= \frac{\Cs^2 - \Si^2}{\Cs^2\,\Si^2}\, A_{n-1}^{\text{Yang-Mills}}(p,3,\ldots,n)\, .
\ee
We have checked this identity independently with the help of explicit four and five point gluon amplitudes expressed via momenta and polarizations presented in \cite{Bjerrum-Bohr:2016juj}. 
It remains to be seen how this intriguingly simple relation may be used to understand the
sub-collinear structure.
This relation arises from  \eqn{eq:gluonkernel} and \eqn{eq:fullfinal}
as 
\be
p\cdot\partial_{\epsilon^{h}_{p}} C_{pp}^{h}=\mathcal{P}_{1}\sim 0\, , \qquad
p\cdot\partial_{\epsilon^{h}_{p}} C_{pp}^{h,(2)}=\mathcal{P}_{2}\, , \qquad
p\cdot\partial_{\epsilon^{h}_{p}} \mathcal{I}_{n-1}^{\text{Yang-Mills}}=0\, ,
\ee
where the first expression vanishes on the support of the scattering equations.

\subsection{Scalars}
The following terms are contributing to the sub-leading order of the collinear limit of scalars in \eqref{eq:chyallintegrand}
\bal{
	\mathcal A_n^{(1)}=2\sum_{\xi_1}\int \underbrace{d\mu_{n-1}^{(0)} \, \mathcal J_1\, \frac{1}{\xi_1^2}\,\mathfrak  C_{n-1}^2}_{t_1} + \underbrace{d\mu_{n-1}^{(0)}\,\mathcal J_0\,2\,\mathfrak  C_n^{(1)}\,\mathfrak  C_{n-1}\,\frac{-1}{\xi_1}}_{t_2} + d'\sigma_{n-1}\,\frac{J_0}{\xi_1^2}\,\mathfrak  C_{n-1}^2\,\Delta_{n-1}'^{(1)} ~.\label{eq:scalarsub}
}
The last term vanishes due to $\Delta_{n-1}'^{(1)}\propto \xi_1$ and therefore the sum over solutions, $\sum_{\xi_1} \frac{\mathcal J_0}{\xi_1} = 0$ kills this term. The other terms are computed to be
\bal{
	&t_1 = -\frac{\Cs^2-\Si^2}{\Cs\,\Si\,D}\left(\frac{\mathcal R_1\,\mathcal Q_1}{2\,p\cdot r} + \mathcal Q_2\right)\,\mathfrak C_{n-1}^2\,d\mu_{n-1}^{(0)}~,\\
	&t_2 = \frac{\Cs^2-\Si^2}{\Cs\,\Si\,D}\left(\frac{\mathcal R_1\,\mathcal Q_1}{2\,p\cdot r} + \mathcal Q_2\right)\,\mathfrak C_{n-1}^2\,d\mu_{n-1}^{(0)}~.
}
Summing up both terms, we see a remarkable cancellation leading to
\beq
\mathcal A_n^{(1)} = 0~,\label{eq:scalarSL}
\eeq
which is a universal theorem, i.e.~independent of the exchange of external particle types, which we shall proof in the appendix. We can express the result via a collinear scalar kernel and write
\beq
\mathcal K_{\text{coll}}^{\text{scalars}}(\{p_i\},\{\sigma_i\})=0~.\label{eq:scalarkernel}
\eeq
We conclude that there is no sub-leading $\mathcal O(\frac{1}{\smallnumber})$ contribution to the collinear scalar scattering.
\section{More universal structure from soft limits}
  In the previous section we derived the universal factorized structure of the CHY integrand for gluons in YM. Now we will show that one can also find a factorized universal structure in \eqref{eq:finalSubL} but this can only be reached by taking an additional soft limit on top of the collinear momentum $p$. Such limits would in fact be a special kinematic configuration of the double soft limits and hence invariably relates the $n$-particle amplitude to an $(n-2)$-particle amplitude. Soft limits are very well studied in the CHY formalism\cite{Cachazo:2013hca} and the calculation is straightforward. Taking $p\rightarrow \delta\, p$ in the sub-leading result \eqref{eq:finalSubL} with $\delta\to 0$ yields the following expansion
  \bal{
  	\mathcal A_{n}^{(1)} = \frac{1}{4\,\pi\,i}&\int d\mu_{n-2}\oint \frac{d\rho}{\delta f_p}\Bigg(\frac{C_{pp}^h}{\delta\mathcal P_2}\left(\frac{1}{\Cs^2}\,\frac{1}{\sigma_{n\rho}}+\frac{1}{\Si^2}\,\frac{1}{\sigma_{\rho3}}\right) \\&+ \frac{\Cs^2 - \Si^2}{\Cs^2\,\Si^2\,\delta\,\mathcal P_2}\,\left(C_{pp}^{h,(2)}-\frac{C_{pp}^h\mathcal P_3}{\mathcal P_2}\right)\Bigg)\,S_{n,\rho,3}\,\mathfrak  C_{n-2}\,C_{pp}^h\,\Pf'(\Psi_{n-2})
  }
 In order to study this limit we have to integrate out the $ \rho $ variable. No pole at infinity is present and therefore we progress in the standard fashion by deforming the contour to include all poles except the one at $f_p =0$, i.e.~the only possible poles are at $\rho \rightarrow \sigma_k ~,~ k \in \{3,..,n\}$. It is also quickly seen that there are no poles for $k \in \{4,...,n-1\}$ and therefore we only need to focus on $k=3$ and $k=n$. Moreover the factor
  \beq
  C_{pp}^{(2)}-\frac{C_{pp}\mathcal P_3}{\mathcal P_2}~,
  \eeq 
  always vanishes when $\rho$ approaches any $\sigma_k$. Hence the only relevant term in the soft limit is
  \beq
  \frac{C_{pp}^2}{f_p\,\mathcal P_2}\,\left(\frac{1}{\Cs^2}\,\frac{1}{\sigma_{n\rho}}+\frac{1}{\Si^2}\,\frac{1}{\sigma_{\rho3}}\right)\,S_{n,\rho,3}\label{eq:relfac}~,
  \eeq
  since neither $\Pf'(\Psi_{n-2})$ nor $C_{n-2}$ have any $\rho$ dependence left. The first contribution comes from $\rho \rightarrow \sigma_3$. In this limit the factor \eqref{eq:relfac} approaches
  \beq
  \frac{1}{\Si^2}\,\frac{1}{\sigma_{\rho 3}}\,\left(\frac{\varepsilon_p\cdot p_3}{p\cdot p_3}\right)^2~.
  \eeq
  and similarly in the limit $\rho \rightarrow \sigma_n$
  \beq
  \frac{1}{\Cs^2}\,\frac{1}{\sigma_{\rho n}}\,\left(\frac{\varepsilon_p\cdot p_n}{p\cdot p_n}\right)^2~.
  \eeq
  Therefore we only get contributions from single poles and we can easily perform the contour integral, obtaining
  \beq
  \mathcal A_n^{(1), p \rightarrow \delta p} = \frac{1}{\delta^2}\left[\frac{1}{\Cs^2}\,\left(\frac{\varepsilon_p\cdot p_n}{p\cdot p_n}\right)^2+\frac{1}{\Si^2}\,\left(\frac{\varepsilon_p\cdot p_3}{p\cdot p_3}\right)^2\right] \mathcal A_{n-2}(3,...,n)~,
  \eeq
 where the superscript of the amplitude indicates the order in which the limits have been performed. We see a universal factorization of the $ n $ point amplitude to an universal factor and a $(n-2)$-point amplitude in the sub-leading collinear and soft limit.

\section{Recovering the Stieberger-Taylor identities}
Stieberger Taylor identities connect sub-leading adjacent collinear gluon limits of pure YM amplitudes with Einstein-Yang-Mills (EYM) amplitudes\cite{Stieberger:2015kia}. In full generality those relations are rather involved and we will refrain from writing them down here. For $n=5$ the they read
\bal{
	s_{5p}\,\mathcal A^{(1)}(1^h,2^h,3,4,5) - s_{4p}\,\mathcal A^{(1)}(1^h,2^h,3,5,4) = \frac{1}{ \Cs^2}A(P^{hh},3,4,5)~,
	}
	where on the LHS we have the sub-leading collinear limit of gluon amplitudes with momenta $p_{1}$ and $p_{2}$ collinear weighted by the Mandelstam invariants $s_{ij}=2\,p_i\cdot p_j$ and on the RHS we have an EYM amplitude where the collinear gluons are replaced by one graviton. We will refer to such linear combinations as ``Stieberger-Taylor (ST) sums" and denote them by $\sum_{ST}$. Considering our result \eqref{eq:finalSubL}, we notice that the only objects affected by any particle permutation are the Parke-Taylor factor and $\frac{1}{\sigma_{n\rho}}+\frac{1}{\sigma_{\rho 3}}$. All other objects are either sums over all particles which remain invariant under permutations, or the Pfaffian in which a particle permutation is equivalent to the interchange of four rows and columns leaving the Pfaffian untouched. In the ST sum additionally $\frac{1}{\sigma_{\rho 3}}$ is invariant since leg $3$ is never permuted. Hence, if we know how $\mathfrak  C_{n-1}$ and $\frac{1}{\sigma_{n\rho}}\,\mathfrak  C_{n-1}$ behave in the ST sum we can deduce the behavior of the amplitude itself. It turns out that while the ST sum exclusively hits $\mathfrak C_{n-1}$, a zero is produced due to either momentum conservation or the production of a $\mathcal P_1$ which is the scattering equation $\bar f_p =0$. We can see this explicitly in the five point case, i.e. 
	\bal{
		&s_{5p}\,\frac{1}{\sigma_{\rho\,3}\,\sigma_{34}\,\sigma_{45}\,\sigma_{5\rho}} - s_{4p}\,\frac{1}{\sigma_{\rho\,3}\,\sigma_{35}\,\sigma_{54}\,\sigma_{4\rho}}
		= s_{5p}\,\left(\frac{1}{\sigma_{5\rho}}+\frac{1}{\sigma_{\rho3}}\right)\,\mathfrak C_3 + s_{4p}\,\left(\frac{1}{\sigma_{4\rho}}+\frac{1}{\sigma_{\rho3}}\right)\,\mathfrak C_3
		\\&=\frac{1}{\sigma_{\rho3}}\,\left(s_{5p}+s_{4p}+s_{3p}\right)\,\mathfrak C_3= 0~,}
	where we used the scattering eq. $\bar f_p =0$ in the third step and $\mathfrak C_3 = \frac{1}{\sigma_{34}\,\sigma_{45}\,\sigma_{53}}$. This behavior was checked up to $n=7$ and we therefore conjecture
	\beq
	\sum_{ST}\,\mathfrak{C_{n-1}} = 0~,
	\eeq
	which is yet to be proven for the general case. The situation changes if we take the ST sum of $\frac{1}{\sigma_{n\rho}}\,\mathfrak  C_{n-1}$, i.e. 
\bal{
	&s_{5p}\,\frac{1}{\sigma_{5\rho}}\,\frac{1}{\sigma_{\rho\,3}\,\sigma_{34}\,\sigma_{45}\,\sigma_{5\rho}} - s_{4p}\,\frac{1}{\sigma_{4\rho}}\,\frac{1}{\sigma_{\rho\,3}\,\sigma_{35}\,\sigma_{54}\,\sigma_{4\rho}}=
	\\& s_{5p}\,\frac{1}{\sigma_{5\rho}}\left(\frac{1}{\sigma_{5\rho}}+\frac{1}{\sigma_{\rho3}}\right)\,\mathfrak C_3 + s_{4p}\,\,\frac{1}{\sigma_{4\rho}}\left(\frac{1}{\sigma_{4\rho}}+\frac{1}{\sigma_{\rho3}}\right)\,\mathfrak C_3 = 
	\\&\left[\frac{s_{5p}}{\sigma_{\rho5}^2}+\frac{s_{4p}}{\sigma_{\rho4}^2} + \frac{1}{\sigma_{\rho3}}\left(\frac{s_{5p}}{\sigma_{5\rho}}+\frac{s_{4p}}{\sigma_{4\rho}}\right)\right]\,\mathfrak C_3 = \mathcal P_2 \, \mathfrak C_3~,
	}
	where we used $\bar f_p =0$ in the last step. This behavior was also confirmed up to $n=7$ and we again conjecture that 
	\beq
	\sum_{ST}\,\mathcal K_{\text{coll}}^{\text{gluon}}(\{p_i\},\{\sigma_i\})\,\mathfrak C_{n-1}=\frac{1}{\mathcal P_2}\,\sum_{ST}\,\frac{1}{\sigma_{n\rho}}\,\mathfrak C_{n-1} = \mathfrak C_{n-2}~.
	\eeq
	Putting all information together we arrive at the statement
	\beq
	\sum_{ST} \mathcal A_n^{(1)}(1^h,2^h,3,...,n) = -\frac{1}{\sqrt 2\,\Cs^2}\int d\mu_{n-1} \, C_{pp}\,\mathfrak C_{n-2}\, \Pf'\Psi_{n-1} = \frac{1}{\Cs^2}\,\mathcal A_{n-1}(P^{hh},3,...,n)~,\label{eq:STfinal}
	\eeq
	which are the ST relations and where the last equality follows from direct construction according to \eqref{eq:EYMint}.
		\section{Collinear limit for effective field theories in CHY}
		In this section we want to compute the leading collinear contribution in the adjacent case for effective field theories that can be constructed in the CHY formalism by methods explained in \cite{Cachazo:2014xea}. Here we probe the leading collinear structure and also comment on the intriguing insight in \cite{Cachazo:2016njl} regarding extension of certain EFT's under single soft limit.  The latter idea essentially means a factorization of an $n$-point CHY amplitude into a sum of lower point CHY amplitudes but living in a theory with extra interactions with additional matter, typically scalars, than those one has started out with and hence the nomenclature `extension'. In order to progress we need to explain further the building blocks and their collinear behavior. We perform the computation for the NLSM model, the Yang-Mills-Scalar theory and the Einstein-Maxwell-Scalar theory. Collinear amplitudes in other EFT's and even sub-leading effects can be computed as a straightforward generalization but we refrain from discussing them here.
		\subsection{More building blocks}
		For the further analysis we need to introduce two more building blocks.  On the one hand we have the unprimed Pfaffian of a $n\times n$ matrix $\mathcal X_n$ with components
		\beq
		\mathcal X_{i,j} =\begin{cases}
			\frac{1}{\sigma_{ij}} & i\neq j \\
			0 & i=j
		\end{cases}~,
		\eeq
		and on the other hand we have the primed Pfaffian of the submatrix $A_n$ of $\Psi_n$. Now we study the adjacent collinear behavior of both building blocks as we did before in section \ref{BBstudy}.
		\paragraph{The Pfaffian $\Pf\mathcal X_n$.}
		We begin by noting the collinear expansion of the relevant entries of $\mathcal X_n$, i.e.
		\bal{
			&\mathcal X_{12} = -\frac{1}{\smallnumber\,\xi_1} + \frac{\xi_2}{\xi_1^2} + \mathcal O(\smallnumber)~,\\&
			\mathcal X_{1b} = \frac{1}{\sigma_{\rho b}} + \smallnumber\,\frac{\xi_1}{2}\,\frac{1}{\sigma_{\rho b}^2} + \mathcal O(\smallnumber^2)~,\\&
			\mathcal X_{2b}= \frac{1}{\sigma_{\rho b}} - \smallnumber\,\frac{\xi_1}{2}\,\frac{1}{\sigma_{\rho b}^2} + \mathcal O(\smallnumber^2)~.
		} 
		We note a divergent behavior in $\mathcal X_{12}$ and therefore the Pfaffian will expand up to leading order as
		\beq
		\Pf\mathcal X_n = -\frac{(-1)^{n+1}}{\sqrt 2 \,\smallnumber\,\xi_1}\,\Pf\mathcal X_n^{1,2}+\mathcal O(1)~,
		\eeq
		where we expanded along the first line.
		\paragraph{The Pfaffian $\Pf'A_n$.}
		We perform the same manipulation scheme as we did in the $h_1=h_2=h$ case for the matrix $\Psi_n$. The first two lines of the matrix $A_n$ are given in \eqref{eq:mentriesaftexp}. We expand the Pfaffian along the first line and obtain 
		\beq
		\Pf'A_n = \frac{(-1)^{n+1}}{\sqrt 2}\,\smallnumber\,\left[\sum_{i=3}^{n}\,(-1)^i\,\left(\xi_1\,\Cs^2\,\Si^2\,A_{pi}^{(2)} - \Cs\,\Si\,\mathcal Q_i\right)\,\Pf'A_n^{1,i}-\frac{2\,p\cdot r}{\xi_1}\,\Pf'A_n^{12}\right] + \mathcal O(\smallnumber^2) ~.
		\eeq
		\subsection{Non-linear sigma model}
		The scattering amplitudes of this model follow from the CHY integrand
		\beq
		\mathcal I_n = \mathfrak C_n\,\left(\Pf'A_n\right)^2~.
		\eeq
		We immediately note that the leading order is of $\mathcal O(\smallnumber)$ in the adjacent collinear limit. More precisely, we can insert the building blocks and write
		\bal{
			\mathcal A_n^{(0)} &= -\smallnumber\,\sum_{\xi_1}\,\int d\mu_{n-1}\,\frac{\mathcal J_0}{\xi_1}\,\mathfrak C_{n-1}\,\Biggl[\sum_{i,j=3}^{n}\,(-1)^{i+j}\,\Bigl(\xi_1^2\,\Cs^4\,\Si^4\,A_{pi}^{(2)}\,A_{pj}^{(2)}-2\,\Cs^3\,\Si^3\,\xi_1\,\mathcal Q_{1,i}\,A_{pj}^{(2)}\\
			&+\Cs^2\,\Si^2\,\mathcal Q_{1,i}\,\mathcal Q_{1,j}\Bigr)\,\Pf'A_n^{1,i}\,\Pf'A_n^{1,j}+4\,\frac{(p\cdot r)^2}{\xi_1^2}\,\left(\Pf'A_n^{1,2}\right)^2\\
			&-4(p\cdot r)\,\sum_{i=3}^{n}\,(-1)^i\,\left(\Cs^2\,\Si^2\,A_{pi}^{(2)}-\frac{\Cs\,\Si\,\mathcal Q_i}{\xi_1}\right)\Pf'A_n^{1,i}\,\Pf'A_n^{1,2}\Biggr]~,
		} 
		resulting in 
		\bal{	\mathcal A_n^{(0)}=&-\smallnumber\,\int d\mu_{n-1}\,\mathfrak C_{n-1}\,\Biggl[\sum_{i,j=3}^{n}\,(-1)^{i+j}\,\left(\frac{\Cs\,\Si\,\mathcal Q_1\,A_{pi}^{(2)}\,A_{pj
				}^{(2)}}{2\,\mathcal P_2^2} - \frac{\Cs\,\Si\,\mathcal Q_{1,i}\,A_{pj}^{(2)}}{\mathcal P_2}\right)\Pf'A_n^{1,i}\,\Pf'A_n^{1,j}
			\\& + \sum_{i=3}^{n}\,(-1)^{i}\,\Cs\,\Si\,\mathcal Q_{1,i}\,\Pf'A_n^{1,i}\,\Pf'A_n^{1,2} - \frac{\Cs\,\Si\,\mathcal Q_1}{2}\,\left(\Pf'A_n^{1,2}\right)^2\Biggr]~.
		\label{eq:collNLSM}}
		
		\subsection{Yang-Mills-Scalar}
		The integrand is given by
		\beq
		\mathcal I_n = \mathfrak C_n\,\Pf\mathcal X_n \, \Pf'A_n~,
		\eeq
		yielding the collinear behavior
		\bal{\mathcal A_n^{(0)} &= -\frac{\Cs\,\Si}{4\,\smallnumber\,(p\cdot r)}\,\int d\mu_{n-1}\,\Pf'\mathcal X_n^{1,2}\,\mathfrak C_{n-1}\,\left[\sum_{i=3}^{n}\,(-1)^i\,\mathcal Q_{1,i}\,\Pf'A_n^{1,i} - \mathcal Q_1\,\Pf'A_n^{1,2}\right]~,
			\label{eq:collYMS}}
		which apart from scalars and gluons, is the only other theory studied here with a singular collinear limit . 
		\subsection{Einstein-Maxwell-Scalar}
		The integrand is 	
		\beq
		\mathcal I_n = \left(\Pf\mathcal X_n \, \Pf'A_n\right)^2~,
		\eeq
		resulting in the collinear limit
		\bal{\mathcal A_n^{(0)} &= \frac{\Cs^2\,\Si^2}{8}\,\int d\mu_{n-1}\,\left(\Pf'\mathcal X_n^{1,2}\right)^2\,\Biggl[\sum_{i,j=3}^{n}\,(-1)^{i+j}\,\left(\frac{2\,A_{pi}^{(2)}\,A_{pj}^{(2)}}{\mathcal P_2} + \frac{\mathcal Q_{1,i}\,\mathcal Q_{1,j}}{(p\cdot r)}\right)\,\Pf'A_n^{1,i}\,\Pf'A_n^{1,j}
			\\&-\frac{2\,\mathcal Q_1}{(p\cdot r)}\,\sum_{i=3}^{n}\,(-1)^i\,\mathcal Q_{1,i}\,\Pf'A_n^{1,i}\,\Pf'A_n^{1,2} + \frac{\mathcal Q_1 +2(p\cdot r)\,\mathcal P_2}{(p\cdot r)}\left(\Pf'A_n^{1,2}\right)^2\Biggr]~.
	\label{eq:collEMS}	}
	
	In all three previous cases we studied in \eqref{eq:collNLSM},\eqref{eq:collYMS}, \eqref{eq:collEMS}, namely NLSM, YMS and EMS the leading collinear behavior is not in a form from which the extension of the theory is manifestly evident and it needs more detailed study to have a conclusive statement regarding this.
\section{Conclusions and Outlook}
		
In this work we initiated the study of collinear limits in massless bosonic quantum field theories beyond the leading order. We realized the CHY formalism is the most appropriate for this purpose. Using the solutions of scattering equations in the collinear kinematics we constructed the perturbative expansion of the CHY integrand building blocks relevant for various field theories. The above expansion allowed us to reproduce all known results for the leading order collinear limit as well as present explicit structures of both the sub-leading collinear limit in pure gauge theories and in $\phi^3$ scalar theories where the latter case in fact has a universal vanishing behavior \eqref{eq:scalarSL}. In particular we carried out a rigorous analysis for the case of adjacent collinear gluons in YM theory. We found that even though the full amplitude does not have a factorized form of a sub-leading splitting function times a lower point amplitude, nevertheless there is a universal factorization at the integrand level of the CHY formula at the sub-leading order. In fact, it is possible to write the CHY integrand of the sub-leading collinear YM amplitude in terms of a lower point YM integrand and a collinear kernel building block \eqref{eq:fullfinal} which is invariant under the exchange of all external particles except the collinear ones and their neighbors and hence results in its universality.

 It is imperative to note that absence of a non-factorized sub-leading splitting function at the integrated amplitude level, is due to the presence of a factor $\mathcal P_2$ in the denominator of the collinear kernel
\eqref{eq:gluonkernel}. This feature is ubiquitous at the sub-leading collinear limit of other theories too and for some special cases even at the leading order. As an example, we also considered the leading collinear behavior of gravitons \eqref{eq:gravL}: Apart from the usual non-singular factorized term, there is also a well-known non-factorized part which can again be attributed to the presence of the same $\mathcal P_2$ term in the denominator of the CHY integrand similar to the sub-leading term in YM theory. Nevertheless, in a remarkable way the above mentioned non-factorized sub-leading collinear amplitude in YM combines through a specific linear combination proposed by Stieberger and Taylor to give amplitudes with one graviton and remaining gluons in EYM theory \eqref{eq:STfinal}. Hence we have been able to directly establish these amplitude relations in field theory which were elusive otherwise.

In summary, we developed a proper treatment of collinear limits in the CHY framework for all integrand  building blocks and hence theories that can be represented by CHY formulas. Treating collinear limits in CHY turns out to be elegant and controlled compared to Feynman graph analysis. It is straightforward to consider even higher order collinear limits or study other EFT's aside from those that we presented here. The collinear limit of the amplitudes in the EFT's studied here, namely for NLSM, YMS and EMS, could not conclusively throw a light on the extension of theories as was seen in single soft limit study of amplitudes in certain EFT's\cite{Cachazo:2016njl}. An important question still needs better understanding though for these purposes, namely the proper treatment of the non-degenerate solutions to $\xi$ with an analytic proof that they never contribute to the sub-leading order which we could only see numerically.

There are few open directions in which one could use the framework for studying collinear effects via CHY formulas. Collinear and soft properties of scattering amplitudes are very important for studying signals in collider experiments. In such a scenario the soft and collinear effects of an amplitude are often bundled in the so called antenna function \cite{Kosower:2003cz}, which is again a complicated object. Now that CHY stands out as the right framework to deal with both soft and collinear effects it will be interesting to have a deeper understanding of antenna functions using CHY. There has also been great progress in understanding scattering amplitudes in $\mathcal{N}=4$ super Yang-Mills theory starting from a near collinear kinematic regime and reconstructing the whole amplitude at any value of coupling \cite{Basso:2014jfa}. This is the OPE approach and it will be interesting to incorporate the ideas of OPE approach into the CHY results for collinear YM\footnote{We thank Lance Dixon for bringing this to our notice.}. Moreover, there has been recent progress in formulating loop amplitudes using CHY formulas (e.g.\cite{Geyer:2015bja,He:2015wgf,He:2015yua,Cachazo:2015aol}) and it will be a natural extension to study soft and collinear limits at the loop level. Finally, we have studied only a few EFTs under the collinear limit and it would definitely be a very important direction to further understand the space of EFTs using soft and collinear limits.
	\section*{Acknowledgments}
We are grateful to Jake Bourjaily, Marco Chiodaroli, Song He, Lance Dixon, Gregory Korchemsky, David Kosower, Oliver Schlotterer, Ellis Yuan and Yang Zhang for various insightful discussions. DN and JP would like to acknowledge the hospitality of Fermilab for the workshop ``MHV@30" and Nordita  for ``Holography and Dualities 2016" and ``Aspects of Amplitudes" during which some of these work had been  done. This work was supported in parts by SFB project ``Space-Time and Matter".
\newpage
\appendix
\section{Sub-leading order calculation}
\subsection{Gluons in pure Yang-Mills}
We divide the calculation into four terms, i.e.
\bal{
	\mathcal A_{n}^{(1)} =2\,\sum_{\xi_1} &\int d'\sigma_{n-2}\,d\rho\Big[\underbrace{\Delta_{n-1}^{(0)} \mathfrak  C_{n}^{(1)}\mathcal J_0 \Pf'^{(0)}\Psi_{n}}_{T_1} +\,\underbrace{\Delta_{n-1}^{(0)} \mathfrak  C_{n}^{(0)}\mathcal J_1 \Pf'^{(0)}\Psi_{n}}_{T_2} \\&+ \underbrace{\Delta_{n-1}^{(0)} \mathfrak  C_{n}^{(0)}\mathcal J_0 \Pf'^{(1)}\Psi_{n}}_{T_3} + \,\underbrace{\Delta_{n-1}^{(1)} \mathfrak  C_{n}^{(0)}\mathcal J_0 \Pf'^{(0)}\Psi_{n}}_{T_4}\Big]~.
	}
The first two terms are easily obtained by using the sum-over-solution relations \eqref{eq:data1} and \eqref{eq:data2}. We will give the relevant structure of the terms which can be adjusted to whether one works in the equal helicity case or the mixed helicity case.

\paragraph{$T_1$: Parke-Taylor contribution.} The relevant term to compute is 
\bal{
	\sum_{\xi_1}\,\mathcal J_0 \,\left(\frac{\xi_2}{\xi_1^2} + \frac{1}{2}\,S_{n,\rho,3}\right)\,\left(\,a\cdot C_{pp}^{h_p} - b\cdot \frac{2\,\epr^{h_p}\cdot p}{\xi_1}\right)~,
	}
where the factors $a$ and $b$ can be adjusted depending on the helicity situation of the collinear gluons, i.e.~whether one uses \eqref{eq:equalpf0} or \eqref{eq:mixpf0} for $\Pf'^{(0)}\Psi_n$. For $h_1=h_2=h$ we have $a=1$ and $b=\frac{1}{\Si\,\Cs}$ and for $h_1\neq h_2$ this term appears twice with either $a=\Si^2$ and $b=\frac{\Si^3}{\Cs}$ or $a=\Cs^2$ and $b=\frac{\Cs^3}{\Si}$. We evaluate this term to be
\bal{
	&a\cdot\frac{(\Cs^s-\Si^2)\,C_{pp}^{h_p}}{4\,\Cs^2\,\Si^2\,D}\,\left(4\,\mathcal R_1 - \frac{\mathcal Q_1\,\mathcal Q_2}{\mathcal P_2}\right) + b\cdot\frac{(\Cs^2-\Si^2)\,\epr^{h_p}\cdot p}{\Cs\,\Si\,D}\,\left(\mathcal Q_2 + \frac{\mathcal Q_1\,\mathcal R_1}{2\,p\cdot r}\right)
	\\&+a\cdot\frac{S_{n,\rho,3}\,C_{pp}^{h_p}}{4\,\Cs^2\,\Si^2\,\mathcal P_2}~.
	}
\paragraph{$T_2$: Jacobian contribution.} The relevant term is
	\bal{
		\sum_{\xi_1}\,\frac{-1}{\xi_1}\,\mJ_0^2\left(8\pr\frac{\xi_2}{\xi_1^3}-\Cs\Si(\Cs^2-\Si^2)\mQ_2\right)\,\left(\,a\cdot C_{pp}^{h_p} - b\cdot \frac{2\,\epr^{h_p}\cdot p}{\xi_1}\right)~,
		}
	which can be evaluated to be 
		\bal{
			-a\cdot\frac{(\Cs^2-\Si^2)\,C_{pp}^{h_p}}{4\,\Cs^2\,\Si^2\,D}\,\left(4\,\mathcal R_1-\frac{\mathcal Q_1 \,\mathcal Q_2}{\mathcal P_2}\right) -b\cdot\frac{(\Cs^2-\Si^2)\,\epr^{h_p}\cdot p}{\Cs\,\Si\,D}\,\left(2\mathcal Q_2 +\frac{\mathcal Q_1\,\mathcal R_1}{p\cdot r}\right)~.
			}
\paragraph{$T_3$: Pfaffian contribution.}
Here we consider the sub-leading expansion of $\Pf'\Psi_n$ for an arbitrary helicity configuration of the collinear gluons after we performed any of the presented manipulation schemes, from either the equal helicity case or the mixed helicity case, and expanded along the first line. Then
\bal{
\Pf'^{(1)}\Psi_{n} &=\frac{-1}{\sqrt 2}\,\frac{\partial}{\partial\smallnumber}\,(x\cdot\Si^2\,C_{11}^{h_1}- y\cdot\Cs^2\,C_{12}^{h_1})\Big|_{\smallnumber=0}\,\Pf'\Psi_{n}^{(1,n+1)} \\&- \frac{1}{\sqrt 2}\,\frac{\partial}{\partial\smallnumber}\,(x\cdot\Cs^2\, C_{22}^{h_2} - y\cdot\Si^2\, C_{21}^{h_2})\Big|_{\smallnumber=0}\,\Pf'\Psi_{n}^{(1,n+2)}
\\&-\frac{1}{\sqrt 2}\,(x\cdot\Si^2\,C_{11}^{h_1}- y\cdot\Cs^2\,C_{12}^{h_1})\,\frac{\partial}{\partial\smallnumber}\,\Pf'\Psi_{n}^{(1,n+1)}\Big|_{\smallnumber=0}   \\&-\frac{1}{\sqrt 2}\,(x\cdot\Cs^2\, C_{22}^{h_2} - y\cdot\Si^2\, C_{21}^{h_2})\,\frac{\partial}{\partial\smallnumber}\,\Pf'\Psi_{n}^{(1,n+2)}\Big|_{\smallnumber=0}~,\label{107}
}
where we introduced the two numbers $x,~y$ in order to quickly switch the manipulation scheme, i.e.~$x=y=1$ for $h_1=h_2=h$ and $x=1,~y=0$ for $h_1\neq h_2$. These are the only relevant terms of any sub-leading Pfaffian expansion of both \eqref{Psiman} and \eqref{Psimanmix} due to the equality of the lines $(n +1)$ and $(n+2)$ in the $h_1 = h_2 =h$ case or due to neglecting any contribution from $\frac{\partial}{\partial\,\smallnumber}\,\mathcal G$ (the reason is an involvement of higher orders in the $\smallnumber$ expansion, i.e.~$\xi = \smallnumber\,\xi_1 + \smallnumber^2\,\xi_2 +\smallnumber^3\,\xi_3+\mathcal O(\smallnumber^4)$, which is beyond the scope of the current discussion) in the mixed helicity case. Let us define the $(2n-2)\times (2n-2)$ matrices $\Psi_n^{(1,n+1)} := \tilde{\Psi}$ and $\Psi_n^{(1,n+2)} := \hat{\Psi}$ for the sake of a compact notation.\\
The contribution to the sub-leading order of the first two terms in \eqref{107} are quickly evaluated to be 
\bal{
	&\sum_{\xi_1} \frac{-\mathcal J_0}{\xi_1}\frac{\partial}{\partial\smallnumber}\,(x\cdot\Si^2\,C_{11}^{h_1}- y\cdot\Cs^2\,C_{12}^{h_1})\Big|_{\smallnumber=0}\,=
	\\&=-\left(x\cdot \Si^2 \frac{C_{pp}^{h_1,(2)}}{4\,\Cs^2\Si^2\,\mathcal P_2} - (x\cdot \Si^2 + y\cdot \Cs^2)\frac{\Si}{\Cs}\,\epr^{h_1}\cdot p \frac{(\Cs^2-\Si^2)}{\Si\,\Cs\,D}\left(\mathcal Q_2 + \frac{\mathcal Q_1\,\mathcal R_1}{2\,p\cdot r}\right)\right)
	} 
	\bal{
		&\sum_{\xi_1} \frac{-\mathcal J_0}{\xi_1}\frac{\partial}{\partial\smallnumber}\,(x\cdot\Cs^2\,C_{22}^{h_2}- y\cdot\Si^2\,C_{21}^{h_2})\Big|_{\smallnumber=0}\,=
		\\&=\left(x\cdot \Cs^2 \frac{C_{pp}^{h_1,(2)}}{4\,\Cs^2\Si^2\,\mathcal P_2} + (x\cdot \Cs^2 + y\cdot \Si^2)\frac{\Cs}{\Si}\,\epr^{h_2}\cdot p \frac{(\Cs^2-\Si^2)}{\Si\,\Cs\,D}\left(\mathcal Q_2 + \frac{\mathcal Q_1\,\mathcal R_1}{2\,p\cdot r}\right)\right)~.
	}
For the last two terms we directly apply \eqref{eq:PfaffianDerviative} without regarding the normalization, i.e. 
\bal{
	\frac{\partial}{\partial\smallnumber}\,\Pf'\tilde\Psi &= \underbrace{\sum_{i=2}^{2\,n-2}\,(-1)^i\,\frac{\partial\,\tilde{\Psi}_{1,i}}{\partial \,\smallnumber}\,\Pf'\tilde{\Psi}^{1,i}}_{\text{Terms with $\smallnumber$ dependence in first row}} + \underbrace{\sum_{i=2}^{n-1}\,(-1)^{n+i+1} \,\frac{\partial\,\tilde{\Psi}_{i,n}}{\partial \,\smallnumber}\,\Pf'\tilde{\Psi}^{i,n}}_{\text{Terms with $\smallnumber$ dependence in $n'th$ column}}
	\\& + \underbrace{\sum_{i=n+1}^{2n-2}\,(-1)^{n+i+1} \,\frac{\partial\,\tilde{\Psi}_{n,i}}{\partial \,\smallnumber}\,\Pf'\tilde{\Psi}^{n,i}}_{\text{Terms with $\smallnumber$ dependence in $n'th$ row}}
	+ \underbrace{\sum_{i=n+1}^{2n-2}\,(-1)^{n} \,\frac{\partial\,\tilde{\Psi}_{n-i+3,i}}{\partial \,\smallnumber}\,\Pf'\tilde{\Psi}^{n-i+3,i}}_{\text{Terms with $\smallnumber$ dependence in the diagonal of $C$}\atop {\text{not including $C_{11}$}}}~.\label{eq:Pfaffianderterms}
	}
The same formula holds true for $\hat \Psi$. We identify
\bal{
	&\tilde{\Psi}_{1,i} = \begin{cases} 
		\tilde A_{2,i+1} & i \leq n-1 \\
		-\tilde C_{i-n+2,2}^{h_2} & i \geq n+1 \\
		-C_{22}^{h_2}-C_{21}^{h_2} & i = n
	\end{cases} \qquad , \qquad \hat{\Psi}_{1,i} = \begin{cases} 
\tilde 	A_{2,i+1} & i \leq n-1 \\
	-\tilde C_{i-n+2,1}^{h_1} & i \geq n+1 \\
	-C_{11}^{h_1}-C_{12}^{h_1} & i = n
\end{cases}~,
	\\&\tilde{\Psi}_{i,n} = -C_{2,i+1}^{h_2} \qquad, \qquad \tilde\Psi_{n,i} = B_{2,i-n+2}^{h_2|h_{i-n+2}}\qquad, \qquad \tilde{\Psi}_{n-i+3,i}= - C_{i-n+2,i-n+2}^{h_{i-n+2}}~,
	\\&\hat{\Psi}_{i,n} = -C_{1,i+1}^{h_1} \qquad, \qquad \hat\Psi_{n,i} = B_{1,i-n+2}^{h_1|h_{i-n+2}}\qquad, \qquad \hat{\Psi}_{n-i+3,i}= - C_{i-n+2,i-n+2}^{h_{i-n+2}}~.
	} 
The rest of the calculation is straightforward. Taking into account the $\xi_1$ dependence of $\mathfrak  C_{n}^{(0)}$ and $\mathcal J_0$, we compute 
\bal{
	&\frac{-1}{\sqrt 2}\,\sum_{\xi_1}\,\frac{-\mathcal J_0}{\xi_1}\,(x\cdot\Si^2\,C_{11}^{h_1}- y\cdot\Cs^2\,C_{12}^{h_1})\,\frac{\partial}{\partial\smallnumber}\,\Pf'\Psi_{n}^{(1,n+1)}\Big|_{\smallnumber=0}=\\ 
	& \frac{1}{\sqrt 2}\,\,\frac{x\cdot\Si^2\,C_{pp}^{h_1}}{4\,\Cs^s\,\Si^2\,\mathcal P_2}\Big\{\sum_{i=2}^{n-1}\,(-1)^i\,(\Cs^2-\Si^2)\,A_{p,i+1}^{(2)}\,\Pf'\tilde{\Psi}^{1,i} \\&+\sum_{i=n+1}^{2n-2}\,(-1)^i\,(\Cs^2-\Si^2)\,C_{i-n+2,p}^{h_{i-n+2},(2)}\,\Pf'\tilde{\Psi}^{1,i}+(-1)^n\,C_{pp}^{h_2,(2)}\,\Pf'\tilde{\Psi}^{1,n}
	\\&+ \sum_{i=n+1}^{2n-2}\,(-1)^{n+i}\,B_{p,i-n+2}^{h_2|h_{i-n+2},(2)}\,\Pf'\tilde{\Psi}^{n,i} - \sum_{i=2}^{n-1}\,(-1)^{n+i}\,C_{p,i+1}^{h_2,(2)}\,\Pf'\tilde{\Psi}^{n,i}
	\\&-\sum_{i=n+1}^{2n-2}\,(-1)^n\,(\Cs^2-\Si^2)\,C_{i-n+2,p}^{h_{i-n+2},(2)}\,\Pf'\tilde{\Psi}^{n-i+3,i}\Big\}
	\\& +\frac{(-1)^{n+1}}{\sqrt 2}\,2\,(x\cdot \Si^2 + y\cdot \Cs^2)\,\epr^{h_1}\cdot p\,E^{h_2}\,\Pf'\tilde{\Psi}^{1,n}~\quad\text{and}\label{eq:psitild}
	}
	\bal{
		&\frac{-1}{\sqrt 2}\,\sum_{\xi_1}\,\frac{-\mathcal J_0}{\xi_1}\,(x\cdot\Cs^2\,C_{22}^{h_2}- y\cdot\Si^2\,C_{21}^{h_2})\,\frac{\partial}{\partial\smallnumber}\,\Pf'\Psi_{n}^{(1,n+2)}\Big|_{\smallnumber=0}=\\ 
		& \frac{1}{\sqrt 2}\,\frac{x\cdot\Cs^2\,C_{pp}^{h_2}}{4\,\Cs^s\,\Si^2\,\mathcal P_2}\Big\{\sum_{i=2}^{n-1}\,(-1)^i\,(\Cs^2-\Si^2)\,A_{p,i+1}^{(2)}\,\Pf'\hat{\Psi}^{1,i} \\&+\sum_{i=n+1}^{2n-2}\,(-1)^i\,(\Cs^2-\Si^2)\,C_{i-n+2,p}^{h_{i-n+2},(2)}\,\Pf'\hat{\Psi}^{1,i}-(-1)^n\,C_{pp}^{h_1,(2)}\,\Pf'\hat{\Psi}^{1,n}
		\\&- \sum_{i=n+1}^{2n-2}\,(-1)^{n+i}\,B_{p,i-n+2}^{h_1|h_{i-n+2},(2)}\,\Pf'\hat{\Psi}^{n,i} + \sum_{i=2}^{n-1}\,(-1)^{n+i}\,C_{p,i+1}^{h_1,(2)}\,\Pf'\hat{\Psi}^{n,i}
		\\&-\sum_{i=n+1}^{2n-2}\,(-1)^n\,(\Cs^2-\Si^2)\,C_{i-n+2,p}^{h_{i-n+2},(2)}\,\Pf'\hat{\Psi}^{n-i+3,i}\Big\}
		\\& -\frac{(-1)^{n+1}}{\sqrt 2}\,2\,(x\cdot \Si^2 + y\cdot \Cs^2)\,\epr^{h_2}\cdot p\,E^{h_1}\,\Pf'\tilde{\Psi}^{1,n}~.\label{eq:psihat}
	}
We can compare both terms with 
\bal{
	&\frac{\partial}{\partial\,\rho}\,\Pf'^\pm\Psi_{n-1}	=\sum_{i=2}^{n-1}\,(-1)^{i+1}\,A_{p,i+1}^{(2)}\,\Pf'^\pm\Psi_{n-1}^{1,i} \\&+\sum_{i=n+1}^{2n-2}\,(-1)^{i+1}\,C_{i-n+2,p}^{h_{i-n+2},(2)}\,\Pf'^\pm\Psi_{n-1}^{1,i}+(-1)^n\,C_{pp}^{\pm,(2)}\,\Pf'^\pm\Psi_{n-1}^{1,n}
	\\&+ \sum_{i=n+1}^{2n-2}\,(-1)^{n+i}\,B_{p,i-n+2}^{\pm|h_{i-n+2},(2)}\,\Pf'^\pm\Psi_{n-1}^{n,i} + \sum_{i=2}^{n-1}\,(-1)^{n+i+1}\,C_{p,i+1}^{\pm,(2)}\,\Pf'^\pm\Psi_{n-1}^{n,i}
	\\&+\sum_{i=n+1}^{2n-2}\,(-1)^n\,C_{i-n+2,p}^{h_{i-n+2},(2)}\,\Pf'^\pm\Psi_{n-1}^{n-i+3,i}~,
	}
to see that summing \eqref{eq:psitild} and \eqref{eq:psihat} in the case of $h_1=h_2=h$ ($x=y=1$) yields
\bal{
	&\quad\frac{-1}{\sqrt 2}\,\sum_{\xi_1}\,\frac{-\mathcal J_0}{\xi_1}\,(\Si^2\,C_{11}^{h}- \Cs^2\,C_{12}^{h})\,\frac{\partial}{\partial\smallnumber}\,\Pf'\Psi_{n}^{(1,n+1)}\Big|_{\smallnumber=0}\\&-\frac{1}{\sqrt 2}\,\sum_{\xi_1}\,\frac{-\mathcal J_0}{\xi_1}\,(\Cs^2\,C_{22}^{h}- \Cs^2\,C_{21}^{h})\,\frac{\partial}{\partial\smallnumber}\,\Pf'\Psi_{n}^{(1,n+2)}\Big|_{\smallnumber=0} \\&= \frac{-1}{\sqrt 2}\,\frac{(\Cs^2-\Si^2)\,C_{pp}^h}{4\,\Cs^2\,\Si^2\,\mathcal P_2}\,\frac{\partial}{\partial\,\rho}\,\Pf'^\pm\Psi_{n-1}~.\label{eq:pfaffder}
	}
\paragraph{$T_4$: $\delta$-functions contribution.}
The relevant term is
\bal{\sum_{\xi_1}\,\frac{-\mathcal J_0}{\xi_1}\,d\,\mu_{n-1}^{(1)}\,\left(\,a\cdot C_{pp}^{h_p} - b\cdot \frac{2\,\epr^{h_p}\cdot p}{\xi_1}\right) =d'\sigma_{n-1}\,\frac{a\cdot (\Cs^2-\Si^2)\,C_{pp}^{h_p}}{4\,\Cs^2\,\Si^2\,\mathcal P_2}\,\frac{\partial}{\partial\,\rho}\,\Delta_{n-1}'~,
	}
which can be quickly seen by direct computation. Now we can sum all contributions while regarding the proper multiplicative factors and quickly arrive at \eqref{eq:subl1} in the $h_1=h_2=h$ case. The case for $h_1\neq h_2$ can also be studied by putting in the proper values of $a,b,x,y$ but the result will not have the feature \eqref{eq:pfaffder}.
\subsection{Gluons $ \rightarrow $  gravitons or scalars}
In the next two sections are heavily based on \cite{Cachazo:2014nsa}. We focus on the EYM integrand given in \eqref{eq:EYMint} with $k$ gravitons and $n-k$ gluons with collinear gluons $1,2$ and particles $3,n$ being gluons s.t. the structure \eqref{PT-linear} is not changed. The sub-leading order of the amplitude is computed via
\bal{
	\mathcal A_{n}^{(1)} &=2\,\sum_{\xi_1} \int d'\sigma_{n-2}\,d\rho\Big[\underbrace{\Delta_{n-1}^{(0)} \mathfrak  C_{n-k}^{(1)}\mathcal J_0\, \Pf^{(0)}\Psi_k\,\Pf'^{(0)}\Psi_{n}}_{T_1} \\&+\,\underbrace{\Delta_{n-1}^{(0)} \mathfrak  C_{n-k}^{(0)}\mathcal J_1 \, \Pf^{(0)}\Psi_k\,\Pf'^{(0)}\Psi_{n}}_{T_2} + \underbrace{\Delta_{n-1}^{(0)} \mathfrak  C_{n-k}^{(0)}\mathcal J_0 \, \Pf^{(0)}\Psi_k\,\Pf'^{(1)}\Psi_{n}}_{T_3} \\&+ \,\underbrace{\Delta_{n-1}^{(1)} \mathfrak  C_{n-k}^{(0)}\mathcal J_0 \, \Pf^{(0)}\Psi_k\,\Pf'^{(0)}\Psi_{n}}_{T_4}+ \underbrace{\Delta_{n-1}^{(0)} \mathfrak  C_{n-k}^{(0)}\mathcal J_0 \, \Pf^{(1)}\Psi_k\,\Pf'^{(0)}\Psi_{n}}_{T_5}\Big]~.
}  
We claim that this yields \eqref{eq:finalSubL} for $h_1=h_2=h$ with the replacements $\mathfrak C_{n-1} \rightarrow \mathfrak C_{n-k-1}$ and $\Pf'\Psi_{n-1} \rightarrow \Pf\Psi_k\,\Pf'\Psi_{n-1}$. In order to prove this claim we only need to show that the last term can be written in terms of a $\rho$ derivative acting on the graviton Pfaffian. This is easy to see, since the only place that the graviton Pfaffian has any $\smallnumber$ dependence are the diagonal entries of the $C$ matrix which is the last term in \eqref{eq:Pfaffianderterms} but this time also including $C_{n-k+1,n-k+1}$, i.e.~the first diagonal element in $C$. Repeating the same computation as in the previous section for $h_1=h_2=h$, we see that 
\beq
\Pf^{(1)}\Psi_k = (\Cs^2-\Si^2)\,\frac{\xi_1}{2}\,\sum_{i=k+1}^{2k}\,C_{n-i+2,p}^{h_{n-i+2},(2)}\,\Pf\Psi_k^{k-i+2,i}~.\label{eq:proof}
\eeq 
If we now compute $T_5$, then we indeed get
\beq
\frac{-1}{\sqrt 2}\,\sum_{\xi_1}\,\frac{-\mathcal J_0}{\xi_1}\,\Pf^{(0)}\Psi_k\,\Pf'^{(0)}\Psi_n = -\frac{1}{\sqrt2}\,\frac{(\Cs^2-\Si^2)\,C_{pp}^h}{4\,\Cs^2\,\Si^2\,\mathcal P_2}\,\Pf'\Psi_{n-1}\,\frac{\partial}{\partial \rho}\,\Pf\Psi_k~.
\eeq
Hence $T_5$ only contributes to the last term in \eqref{eq:finalSubL} and the kernel \eqref{eq:gluonkernel} is unaffected by the exchange of gluons to gravitons as long as the collinear particles and their neighbors are gluons. Changing gluons into scalars also cannot have any effect on the structure of \eqref{eq:finalSubL} since the only thing that will happen is that the size of the matrix $\Psi_n$ will change into $\Psi_{n-k}$ for $k$ scalars and that some parts of $\mathfrak C_n$ will now be squared but those parts have no $\smallnumber$ dependence since particles $1,2,3,n$ are fixed to be gluons. Therefore \eqref{eq:finalSubL} is true up to an adjustment of $\mathfrak C_{n-1}$ and $\Pf'\Psi_{n-1} \rightarrow \Pf'\Psi_{n-k-1}$. Thus is kernel \eqref{eq:gluonkernel} is again invariant, which proves universality.
\subsection{Scalars $ \rightarrow $ gravitons or gluons} 
We can quickly show that \eqref{eq:scalarSL} is also independent of the exchange of all external particles except the collinear scalars and their neighbors. The reason is that if we start exchanging the particle types we start adjusting the size of the Parke-Taylor structure in \eqref{eq:scalarsub}, which is not affecting the established structure, and adding Pfaffians of matrices $\Psi_{k}$ for $k$ gluons or gravitons. The Pfaffian structure will yield more terms to the sub-leading structure but those terms will vanish due to $\sum_{\xi_1}\,\frac{\mathcal J_0}{\xi_1}=0$ since $\mathfrak (C_{n-k}^{(0)})^2 \propto \frac{1}{\xi_1^2}$ and $\Pf^{(1)} \Psi_k \propto \xi_1$ as we have shown in \eqref{eq:proof}. Hence no corrections to the sub-leading order is present and the kernel \eqref{eq:scalarkernel} is invariant and universality is manifest.
\section{Non-adjacent collinear gluons of equal helicity}
  Let us take $p_i || p_j$ for some non-adjacent $i$ and $j$. We repeat all steps that we did in the adjacent case (i.e.~replace $1\rightarrow i$ and $2\rightarrow j$) but with the additional, follow up, matrix manipulation:
  \begin{itemize}
  	\item Subtract $(n+j)$'th row/column from $(n+i)$'th row/column.
  \end{itemize}
  Then the Pfaffian expands along the $i$'th row as (up to leading order) 
  \beq
  \Pf'(\Psi_n)=\frac{1}{\sqrt 2}\,(-1)^{i+i}\,\left(-C_{pp}^h + \frac{2\,\epr^h\cdot p}{\Cs\Si\,\xi_1}\right)\Pf'(\Psi_{n-1})~.
  \eeq
  Since there is no divergence in the non-adjacent case (i.e.~there is no divergence coming from the Parke-Taylor factor) the amplitude expands as
  \beq
  \mathcal A_n = \frac{1}{\sqrt 2}\,\sum_{\xi_1}\,2\,\int d\mu_{n-1}\,\mJ_0\,\left(C_{pp}^h - \frac{2\,\epr\cdot p}{\Cs\Si\,\xi_1}\right)\, S_{i-1,\rho,i+1}\,S_{j-1,\rho,j+1}\,\mathfrak  C_{n-2}\,\Pf'(\Psi_{n-1}) + \mathcal O(\epsilon)~,
  \eeq
  which results in
  \beq
  \mathcal A_n = \frac{1}{\sqrt 2\,\Cs^2\,\Si^2}\int d\mu_{n-1}\,\frac{C_{pp}^h}{\mathcal P_2} S_{i-1,\rho,i+1}\,S_{j-1,\rho,j+1}\,\mathfrak  C_{n-2}\,\Pf'(\Psi_{n-1})~.
  \eeq
  The indices of the factors $S_{k,l,m}$ refer to the position of $i$ and $j$ before the collinear limit and $C_{n-2}$ does not contain any $\rho$ label. We can also check that this result reproduces the known $ST$ relations \cite{Stieberger:2015kia}. Nevertheless, although universal, no factorization at amplitude level can be extracted.
\newpage

\bibliographystyle{nb}
\bibliography{soft}

\begin{thebibliography}{10}
\providecommand{\href}[2]{#2}
\providecommand{\arxivref}[2]{\href{http://arxiv.org/abs/#1}{#2}}
\providecommand{\doiref}[2]{\href{http://dx.doi.org/#1}{#2}}
\providecommand{\nbbstauthor}[1]{#1}
\providecommand{\nbbstjournal}[1]{\textsf{#1}}
\providecommand{\nbbsttitle}[1]{\textit{#1}}
\providecommand{\nbbsturl}[1]{\texttt{#1}}
\providecommand{\nbbsteprint}[1]{\texttt{#1}}
\providecommand{\nbbststyle}{\raggedright\small\parskip0pt}
\nbbststyle

\bibitem{Dixon:1996wi}
\nbbstauthor{L.~J.~Dixon},
\nbbsttitle{``{Calculating scattering amplitudes efficiently}''},
\nbbsteprint{\arxivref{hep-ph/9601359}{hep-ph/9601359}}.

\bibitem{Elvang:2013cua}
\nbbstauthor{H.~Elvang and Y.-t.~Huang},
\nbbsttitle{``{Scattering Amplitudes in Gauge Theory and Gravity}''},
\nbbstjournal{Cambridge~University~Press~,~~(2015)},
\nbbsteprint{\arxivref{1308.1697}{arxiv:1308.1697}}.

\bibitem{Henn:2014yza}
\nbbstauthor{J.~M.~Henn and J.~C.~Plefka},
\nbbsttitle{``{Scattering Amplitudes in Gauge Theories}''},
\nbbstjournal{\doiref{978-3-642-54021-9,
  10.1007/978-3-642-54022-6}{Lect.~Notes~Phys.~883,~1~(2014)}}.

\bibitem{Weinberg:1964ew}
\nbbstauthor{S.~Weinberg},
\nbbsttitle{``{Photons and Gravitons in s Matrix Theory: Derivation of Charge
  Conservation and Equality of Gravitational and Inertial Mass}''},
\nbbstjournal{\doiref{10.1103/PhysRev.135.B1049}{Phys.~Rev.~135,~B1049~(1964)}}.

\bibitem{Low:1958sn}
\nbbstauthor{F.~Low},
\nbbsttitle{``{Bremsstrahlung of very low-energy quanta in elementary particle
  collisions}''},
\nbbstjournal{\doiref{10.1103/PhysRev.110.974}{Phys.~Rev.~110,~974~(1958)}}.

\bibitem{Burnett:1967km}
\nbbstauthor{T.~Burnett and N.~M.~Kroll},
\nbbsttitle{``{Extension of the low soft photon theorem}''},
\nbbstjournal{\doiref{10.1103/PhysRevLett.20.86}{Phys.~Rev.~Lett.~20,~86~(1968)}}.

\bibitem{Casali:2014xpa}
\nbbstauthor{E.~Casali},
\nbbsttitle{``{Soft sub-leading divergences in Yang-Mills amplitudes}''},
\nbbstjournal{\doiref{10.1007/JHEP08(2014)077}{JHEP~1408,~077~(2014)}},
\nbbsteprint{\arxivref{1404.5551}{arxiv:1404.5551}}.

\bibitem{Cachazo:2014fwa}
\nbbstauthor{F.~Cachazo and A.~Strominger},
\nbbsttitle{``{Evidence for a New Soft Graviton Theorem}''},
\nbbsteprint{\arxivref{1404.4091}{arxiv:1404.4091}}.

\bibitem{Bern:2005iz}
\nbbstauthor{Z.~Bern, L.~J.~Dixon and V.~A.~Smirnov},
\nbbsttitle{``{Iteration of planar amplitudes in maximally supersymmetric
  Yang-Mills theory at three loops and beyond}''},
\nbbstjournal{\doiref{10.1103/PhysRevD.72.085001}{Phys.~Rev.~D72,~085001~(2005)}},
\nbbsteprint{\arxivref{hep-th/0505205}{hep-th/0505205}}.

\bibitem{Anastasiou:2003kj}
\nbbstauthor{C.~Anastasiou, Z.~Bern, L.~J.~Dixon and D.~A.~Kosower},
\nbbsttitle{``{Planar amplitudes in maximally supersymmetric Yang-Mills
  theory}''},
\nbbstjournal{\doiref{10.1103/PhysRevLett.91.251602}{Phys.~Rev.~Lett.~91,~251602~(2003)}},
\nbbsteprint{\arxivref{hep-th/0309040}{hep-th/0309040}}.

\bibitem{Strominger:2013jfa}
\nbbstauthor{A.~Strominger},
\nbbsttitle{``{On BMS Invariance of Gravitational Scattering}''},
\nbbsteprint{\arxivref{1312.2229}{arxiv:1312.2229}}.

\bibitem{He:2014laa}
\nbbstauthor{T.~He, V.~Lysov, P.~Mitra and A.~Strominger},
\nbbsttitle{``{BMS supertranslations and Weinberg's soft graviton theorem}''},
\nbbsteprint{\arxivref{1401.7026}{arxiv:1401.7026}}.

\bibitem{He:2015zea}
\nbbstauthor{T.~He, P.~Mitra and A.~Strominger},
\nbbsttitle{``{2D Kac-Moody Symmetry of 4D Yang-Mills Theory}''},
\nbbsteprint{\arxivref{1503.02663}{arxiv:1503.02663}}.

\bibitem{Britto:2004ap}
\nbbstauthor{R.~Britto, F.~Cachazo and B.~Feng},
\nbbsttitle{``{New recursion relations for tree amplitudes of gluons}''},
\nbbstjournal{\doiref{10.1016/j.nuclphysb.2005.02.030}{Nucl.~Phys.~B715,~499~(2005)}},
\nbbsteprint{\arxivref{hep-th/0412308}{hep-th/0412308}}.

\bibitem{Britto:2005fq}
\nbbstauthor{R.~Britto, F.~Cachazo, B.~Feng and E.~Witten},
\nbbsttitle{``{Direct proof of tree-level recursion relation in Yang-Mills
  theory}''},
\nbbstjournal{\doiref{10.1103/PhysRevLett.94.181602}{Phys.~Rev.~Lett.~94,~181602~(2005)}},
\nbbsteprint{\arxivref{hep-th/0501052}{hep-th/0501052}}.

\bibitem{Cachazo:2013hca}
\nbbstauthor{F.~Cachazo, S.~He and E.~Y.~Yuan},
\nbbsttitle{``{Scattering of Massless Particles in Arbitrary Dimensions}''},
\nbbstjournal{\doiref{10.1103/PhysRevLett.113.171601}{Phys.~Rev.~Lett.~113,~171601~(2014)}},
\nbbsteprint{\arxivref{1307.2199}{arxiv:1307.2199}}.

\bibitem{Cachazo:2013iea}
\nbbstauthor{F.~Cachazo, S.~He and E.~Y.~Yuan},
\nbbsttitle{``{Scattering of Massless Particles: Scalars, Gluons and
  Gravitons}''},
\nbbstjournal{\doiref{10.1007/JHEP07(2014)033}{JHEP~1407,~033~(2014)}},
\nbbsteprint{\arxivref{1309.0885}{arxiv:1309.0885}}.

\bibitem{Fairlie:2008dg}
\nbbstauthor{D.~B.~Fairlie},
\nbbsttitle{``{A Coding of Real Null Four-Momenta into World-Sheet
  Co-ordinates}''},
\nbbstjournal{\doiref{10.1155/2009/284689}{Adv.~Math.~Phys.~2009,~284689~(2009)}},
\nbbsteprint{\arxivref{0805.2263}{arxiv:0805.2263}}.

\bibitem{Gross:1987ar}
\nbbstauthor{D.~J.~Gross and P.~F.~Mende},
\nbbsttitle{``{String Theory Beyond the Planck Scale}''},
\nbbstjournal{\doiref{10.1016/0550-3213(88)90390-2}{Nucl.~Phys.~B303,~407~(1988)}}.

\bibitem{Witten:2004cp}
\nbbstauthor{E.~Witten},
\nbbsttitle{``{Parity invariance for strings in twistor space}''},
\nbbstjournal{\doiref{10.4310/ATMP.2004.v8.n5.a1}{Adv.~Theor.~Math.~Phys.~8,~779~(2004)}},
\nbbsteprint{\arxivref{hep-th/0403199}{hep-th/0403199}}.

\bibitem{Caputa:2011zk}
\nbbstauthor{P.~Caputa and S.~Hirano},
\nbbsttitle{``{Observations on Open and Closed String Scattering Amplitudes at
  High Energies}''},
\nbbstjournal{\doiref{10.1007/JHEP02(2012)111}{JHEP~1202,~111~(2012)}},
\nbbsteprint{\arxivref{1108.2381}{arxiv:1108.2381}}.

\bibitem{Caputa:2012pi}
\nbbstauthor{P.~Caputa},
\nbbsttitle{``{Lightlike contours with fermions}''},
\nbbstjournal{\doiref{10.1016/j.physletb.2012.09.006}{Phys.~Lett.~B716,~475~(2012)}},
\nbbsteprint{\arxivref{1205.6369}{arxiv:1205.6369}}.

\bibitem{Makeenko:2011dm}
\nbbstauthor{Y.~Makeenko and P.~Olesen},
\nbbsttitle{``{The QCD scattering amplitude from area behaved Wilson loops}''},
\nbbstjournal{\doiref{10.1016/j.physletb.2012.02.032}{Phys.~Lett.~B709,~285~(2012)}},
\nbbsteprint{\arxivref{1111.5606}{arxiv:1111.5606}}.

\bibitem{Cachazo:2012uq}
\nbbstauthor{F.~Cachazo},
\nbbsttitle{``{Fundamental BCJ Relation in N=4 SYM From The Connected
  Formulation}''},
\nbbsteprint{\arxivref{1206.5970}{arxiv:1206.5970}}.

\bibitem{Cachazo:2015ksa}
\nbbstauthor{F.~Cachazo, S.~He and E.~Y.~Yuan},
\nbbsttitle{``{New Double Soft Emission Theorems}''},
\nbbsteprint{\arxivref{1503.04816}{arxiv:1503.04816}}.

\bibitem{He:2016vfi}
\nbbstauthor{S.~He, Z.~Liu and J.-B.~Wu},
\nbbsttitle{``{Scattering Equations, Twistor-string Formulas and Double-soft
  Limits in Four Dimensions}''},
\nbbstjournal{\doiref{10.1007/JHEP07(2016)060}{JHEP~1607,~060~(2016)}},
\nbbsteprint{\arxivref{1604.02834}{arxiv:1604.02834}}.

\bibitem{Saha:2016kjr}
\nbbstauthor{A.~P.~Saha},
\nbbsttitle{``{Double Soft Theorem for Perturbative Gravity}''},
\nbbsteprint{\arxivref{1607.02700}{arxiv:1607.02700}}.

\bibitem{Klose:2015xoa}
\nbbstauthor{T.~Klose, T.~McLoughlin, D.~Nandan, J.~Plefka and G.~Travaglini},
\nbbsttitle{``{Double-Soft Limits of Gluons and Gravitons}''},
\nbbstjournal{\doiref{10.1007/JHEP07(2015)135}{JHEP~1507,~135~(2015)}},
\nbbsteprint{\arxivref{1504.05558}{arxiv:1504.05558}}.

\bibitem{Volovich:2015yoa}
\nbbstauthor{A.~Volovich, C.~Wen and M.~Zlotnikov},
\nbbsttitle{``{Double Soft Theorems in Gauge and String Theories}''},
\nbbstjournal{\doiref{10.1007/JHEP07(2015)095}{JHEP~1507,~095~(2015)}},
\nbbsteprint{\arxivref{1504.05559}{arxiv:1504.05559}}.

\bibitem{Stieberger:2009hq}
\nbbstauthor{S.~Stieberger},
\nbbsttitle{``{Open \& Closed vs. Pure Open String Disk Amplitudes}''},
\nbbsteprint{\arxivref{0907.2211}{arxiv:0907.2211}}.

\bibitem{Stieberger:2015vya}
\nbbstauthor{S.~Stieberger and T.~R.~Taylor},
\nbbsttitle{``{Disk Scattering of Open and Closed Strings (I)}''},
\nbbstjournal{\doiref{10.1016/j.nuclphysb.2015.12.002}{Nucl.~Phys.~B903,~104~(2016)}},
\nbbsteprint{\arxivref{1510.01774}{arxiv:1510.01774}}.

\bibitem{Stieberger:2016lng}
\nbbstauthor{S.~Stieberger and T.~R.~Taylor},
\nbbsttitle{``{New Relations for Einstein-Yang-Mills Amplitudes}''},
\nbbsteprint{\arxivref{1606.09616}{arxiv:1606.09616}}.

\bibitem{Schlotterer:2016cxa}
\nbbstauthor{O.~Schlotterer},
\nbbsttitle{``{Amplitude relations in heterotic string theory and
  Einstein-Yang-Mills}''},
\nbbsteprint{\arxivref{1608.00130}{arxiv:1608.00130}}.

\bibitem{Nandan:2016pya}
\nbbstauthor{D.~Nandan, J.~Plefka, O.~Schlotterer and C.~Wen},
\nbbsttitle{``{Einstein-Yang-Mills from pure Yang-Mills amplitudes}''},
\nbbsteprint{\arxivref{1607.05701}{arxiv:1607.05701}}.

\bibitem{delaCruz:2016gnm}
\nbbstauthor{L.~de~la~Cruz, A.~Kniss and S.~Weinzierl},
\nbbsttitle{``{Relations for Einstein-Yang-Mills amplitudes from the CHY
  representation}''},
\nbbsteprint{\arxivref{1607.06036}{arxiv:1607.06036}}.

\bibitem{Stieberger:2015kia}
\nbbstauthor{S.~Stieberger and T.~R.~Taylor},
\nbbsttitle{``{Subleading terms in the collinear limit of Yang?Mills
  amplitudes}''},
\nbbstjournal{\doiref{10.1016/j.physletb.2015.09.075}{Phys.~Lett.~B750,~587~(2015)}},
\nbbsteprint{\arxivref{1508.01116}{arxiv:1508.01116}}.

\bibitem{Stieberger:2014cea}
\nbbstauthor{S.~Stieberger and T.~R.~Taylor},
\nbbsttitle{``{Graviton as a Pair of Collinear Gauge Bosons}''},
\nbbstjournal{\doiref{10.1016/j.physletb.2014.10.057}{Phys.~Lett.~B739,~457~(2014)}},
\nbbsteprint{\arxivref{1409.4771}{arxiv:1409.4771}}.

\bibitem{Cachazo:2014xea}
\nbbstauthor{F.~Cachazo, S.~He and E.~Y.~Yuan},
\nbbsttitle{``{Scattering Equations and Matrices: From Einstein To Yang-Mills,
  DBI and NLSM}''},
\nbbstjournal{\doiref{10.1007/JHEP07(2015)149}{JHEP~1507,~149~(2015)}},
\nbbsteprint{\arxivref{1412.3479}{arxiv:1412.3479}}.

\bibitem{Cheung:2014dqa}
\nbbstauthor{C.~Cheung, K.~Kampf, J.~Novotny and J.~Trnka},
\nbbsttitle{``{Effective Field Theories from Soft Limits of Scattering
  Amplitudes}''},
\nbbstjournal{\doiref{10.1103/PhysRevLett.114.221602}{Phys.~Rev.~Lett.~114,~221602~(2015)}},
\nbbsteprint{\arxivref{1412.4095}{arxiv:1412.4095}}.

\bibitem{Cheung:2015ota}
\nbbstauthor{C.~Cheung, K.~Kampf, J.~Novotny, C.-H.~Shen and J.~Trnka},
\nbbsttitle{``{On-Shell Recursion Relations for Effective Field Theories}''},
\nbbstjournal{\doiref{10.1103/PhysRevLett.116.041601}{Phys.~Rev.~Lett.~116,~041601~(2016)}},
\nbbsteprint{\arxivref{1509.03309}{arxiv:1509.03309}}.

\bibitem{Cachazo:2014nsa}
\nbbstauthor{F.~Cachazo, S.~He and E.~Y.~Yuan},
\nbbsttitle{``{Einstein-Yang-Mills Scattering Amplitudes From Scattering
  Equations}''},
\nbbstjournal{\doiref{10.1007/JHEP01(2015)121}{JHEP~1501,~121~(2015)}},
\nbbsteprint{\arxivref{1409.8256}{arxiv:1409.8256}}.

\bibitem{Dolan:2014ega}
\nbbstauthor{L.~Dolan and P.~Goddard},
\nbbsttitle{``{The Polynomial Form of the Scattering Equations}''},
\nbbstjournal{\doiref{10.1007/JHEP07(2014)029}{JHEP~1407,~029~(2014)}},
\nbbsteprint{\arxivref{1402.7374}{arxiv:1402.7374}}.

\bibitem{Bern:2004cz}
\nbbstauthor{Z.~Bern, L.~J.~Dixon and D.~A.~Kosower},
\nbbsttitle{``{Two-loop g ---> gg splitting amplitudes in QCD}''},
\nbbstjournal{\doiref{10.1088/1126-6708/2004/08/012}{JHEP~0408,~012~(2004)}},
\nbbsteprint{\arxivref{hep-ph/0404293}{hep-ph/0404293}}.

\bibitem{Bern:1998xc}
\nbbstauthor{Z.~Bern, L.~J.~Dixon, M.~Perelstein and J.~S.~Rozowsky},
\nbbsttitle{``{One loop n point helicity amplitudes in (selfdual) gravity}''},
\nbbstjournal{\doiref{10.1016/S0370-2693(98)01397-5}{Phys.~Lett.~B444,~273~(1998)}},
\nbbsteprint{\arxivref{hep-th/9809160}{hep-th/9809160}}.

\bibitem{Bern:1998sv}
\nbbstauthor{Z.~Bern, L.~J.~Dixon, M.~Perelstein and J.~S.~Rozowsky},
\nbbsttitle{``{Multileg one loop gravity amplitudes from gauge theory}''},
\nbbstjournal{\doiref{10.1016/S0550-3213(99)00029-2}{Nucl.~Phys.~B546,~423~(1999)}},
\nbbsteprint{\arxivref{hep-th/9811140}{hep-th/9811140}}.

\bibitem{Bjerrum-Bohr:2016juj}
\nbbstauthor{N.~E.~J.~Bjerrum-Bohr, J.~L.~Bourjaily, P.~H.~Damgaard and
  B.~Feng},
\nbbsttitle{``{Analytic Representations of Yang-Mills Amplitudes}''},
\nbbsteprint{\arxivref{1605.06501}{arxiv:1605.06501}}.

\bibitem{Cachazo:2016njl}
\nbbstauthor{F.~Cachazo, P.~Cha and S.~Mizera},
\nbbsttitle{``{Extensions of Theories from Soft Limits}''},
\nbbstjournal{\doiref{10.1007/JHEP06(2016)170}{JHEP~1606,~170~(2016)}},
\nbbsteprint{\arxivref{1604.03893}{arxiv:1604.03893}}.

\bibitem{Kosower:2003cz}
\nbbstauthor{D.~A.~Kosower},
\nbbsttitle{``{All orders singular emission in gauge theories}''},
\nbbstjournal{\doiref{10.1103/PhysRevLett.91.061602}{Phys.~Rev.~Lett.~91,~061602~(2003)}},
\nbbsteprint{\arxivref{hep-ph/0301069}{hep-ph/0301069}}.

\bibitem{Basso:2014jfa}
\nbbstauthor{B.~Basso, A.~Sever and P.~Vieira},
\nbbsttitle{``{Collinear Limit of Scattering Amplitudes at Strong Coupling}''},
\nbbstjournal{\doiref{10.1103/PhysRevLett.113.261604}{Phys.~Rev.~Lett.~113,~261604~(2014)}},
\nbbsteprint{\arxivref{1405.6350}{arxiv:1405.6350}}.

\bibitem{Geyer:2015bja}
\nbbstauthor{Y.~Geyer, L.~Mason, R.~Monteiro and P.~Tourkine},
\nbbsttitle{``{Loop Integrands for Scattering Amplitudes from the Riemann
  Sphere}''},
\nbbstjournal{\doiref{10.1103/PhysRevLett.115.121603}{Phys.~Rev.~Lett.~115,~121603~(2015)}},
\nbbsteprint{\arxivref{1507.00321}{arxiv:1507.00321}}.

\bibitem{He:2015wgf}
\nbbstauthor{S.~He, R.~Monteiro and O.~Schlotterer},
\nbbsttitle{``{String-inspired BCJ numerators for one-loop MHV amplitudes}''},
\nbbstjournal{\doiref{10.1007/JHEP01(2016)171}{JHEP~1601,~171~(2016)}},
\nbbsteprint{\arxivref{1507.06288}{arxiv:1507.06288}}.

\bibitem{He:2015yua}
\nbbstauthor{S.~He and E.~Y.~Yuan},
\nbbsttitle{``{One-loop Scattering Equations and Amplitudes from Forward
  Limit}''},
\nbbstjournal{\doiref{10.1103/PhysRevD.92.105004}{Phys.~Rev.~D92,~105004~(2015)}},
\nbbsteprint{\arxivref{1508.06027}{arxiv:1508.06027}}.

\bibitem{Cachazo:2015aol}
\nbbstauthor{F.~Cachazo, S.~He and E.~Y.~Yuan},
\nbbsttitle{``{One-Loop Corrections from Higher Dimensional Tree
  Amplitudes}''},
\nbbstjournal{\doiref{10.1007/JHEP08(2016)008}{JHEP~1608,~008~(2016)}},
\nbbsteprint{\arxivref{1512.05001}{arxiv:1512.05001}}.

\end{thebibliography}

\end{document}